\documentclass[fleqn,usenatbib]{mnras}

\usepackage{newtxtext,newtxmath}
\usepackage{amsmath,graphicx}
\usepackage{longtable}
\usepackage{supertabular}

\usepackage[T1]{fontenc}

\DeclareRobustCommand{\VAN}[3]{#2}
\let\VANthebibliography\thebibliography
\def\thebibliography{\DeclareRobustCommand{\VAN}[3]{##3}\VANthebibliography}

\usepackage{graphicx}	
\usepackage{amsmath}	
\usepackage{newtxtext,newtxmath}
\usepackage{soul}

\title[GRB 180325A: dust grain-size distribution model]{GRB\,180325A: dust grain-size distribution and interstellar iron nanoparticles contribution}

\author[Cappellazzo et al.]{
Elizabeth Cappellazzo,$^{1,2}$\thanks{E-mail: elizabeth.cappellazzo@students.mq.edu.au}
Tayyaba Zafar,$^{1,2,3}$
Pablo Corcho-Caballero$^{1,3,4}$
David Alexander Kann,$^{5}$
\newauthor
\'Angel L\'opez-S\'anchez,$^{1,2,3}$
and Adeel Ahmad,$^{6}$
\\
$^{1}$Australian Astronomical Optics, Macquarie University, 105 Delhi Road, North Ryde, NSW 2113, Australia\\
$^{2}$ Macquarie University Research Centre for Astronomy, Astrophysics \& Astrophotonics, Sydney, NSW 2109, Australia\\
$^{3}$ARC Centre of Excellence for All Sky Astrophysics in 3 Dimensions (ASTRO-3D), Australia\\
$^{4}$ Departamento de F\'isicia Te\'orica, Universidad Aut\'onoma de Madrid (UAM), Campus de Cantoblanco, E-28049 Madrid, Spain\\
$^{5}$Instituto de Astrof\'isica de Andaluc\'ia (IAA-CSIC), Glorieta de la Astronom\'ia s/n, 18008 Granada, Spain\\
$^{6}$ School of Science, Western Sydney University, Locked Bag 1797, Penrith South DC, NSW 2751, Australia}

\date{Accepted XXX. Received YYY; in original form ZZZ}

\pubyear{2022}

\begin{document}
\label{firstpage}
\pagerange{\pageref{firstpage}--\pageref{lastpage}}
\maketitle

\begin{abstract}
We modelled dust grain-size distributions for carbonaceous and silicates dust, as well as for free-flying iron nanoparticles in the environment of a $\gamma$-ray burst (GRB) afterglow, GRB\,180325A.
This GRB, at $z=2.2486$, has an unambiguous detection of the 2175\,\AA\ extinction feature with $R_V=4.58$ and $A_V=1.58$.
In addition to silicates, polycyclic aromatic hydrocarbons (PAH), and graphite, we used iron nanoparticles grain-size distributions for the first time to model the observed extinction curve of GRB\,180325A.
We fit the observed extinction for four model permutations, using 232 sets of silicates, graphite, carbon abundance in hydrocarbon molecules ($b_C$), and fraction of iron abundance in free-flying nanoparticles ($b_{\text{Fe}}$).
These four different permutations were chosen to test iron nanoparticles significance and carbon abundance in hydrocarbons.
Our results indicate that iron nanoparticles contribution is insignificant and there is a degeneracy of carbon abundances, with the range $(0.0 \leq b_C \leq 0.7)\times10^{-5}$ providing the best-fit to the observed extinction curve of GRB\,180325A.
We therefore favour the simplest model of silicates and polycyclic aromatic hydrocarbons. The silicates are dominant and contribute to the entire wavelength range of the GRB extinction curve while graphite contributes towards both the 2175\,\AA\ bump and the UV extinction.
The afterglow peak luminosity ($1.5\times10^{51}$\,ergs/s) indicates dust destruction may have taken place.
We conclude that further investigations into other potential contributors of extinction are warranted, particularly for steep UV extinction.
\end{abstract}




\section{Introduction}
Extinction is an important and meaningful tool to measure the dust properties in the interstellar medium. Modelling extinction curves in optical light from astronomical objects with a dust grain-size distribution is used to understand the environments that the light passes through, within the Milky Way (MW) Galaxy and beyond. The classic interstellar dust model from \citet{Mathis77} consists of spherical homogeneous graphite and silicates dust grains with a grain-size distribution $dn/da$ $\propto$ $a^{-3.5}$ where $a$ is the grain radius with $a_{\text{min}} = 50$\,\AA\ and $a_{\text{max}} = 0.25$\,$\mu$m. The \citet{Mathis77} model assumed that the light extinction curve is the same for all sightlines. They considered spherical grains and used Mie theory to compute extinction cross sections. Mie scattering theory is predominately used in dust grain models, where dust grains are treated as symmetrical, homogeneous spheres. 

Dust grain models have evolved beyond this as more observational evidence shows that extinction curves vary for different lines of sight. This implies that the environment through which light travels is not homogeneous across lines of sight. \citet{Cardelli89} defined total-to-selective extinction $R_V \equiv A(V)/E(B-V)$ which is the ratio of visual extinction to reddening and can be used as a tool to characterise these differences in sightlines. Many interstellar extinction lines of sight within the Galaxy have been modelled with dust grain-size distributions, as well as lines of sight from the Large and Small Magellanic Clouds \citep[LMC and SMC;][]{Draine84,Zubko98,Weingartner01, Li01, Muthumariappan10, Gao13, Poteet15}.

The 2175\,\AA\ extinction feature was first discovered by \citet{Stecher65}, occurring in all sightlines within the MW, most of the sightlines in the LMC, and in one sightline towards the SMC. It is attributed to extinction from small carbonaceous dust grains \citep{Draine03}. A 2175\,\AA\ bump is very uncommon beyond the local universe \citep[e.g.,][]{Zeimann15,Reddy18}. To explain extinction curves in the local universe of a wide range of environments demonstrating a 2175\,\AA\ feature, many dust grain models containing plentiful very small sized grains have been computed with the so-called `three component' dust grain model. This dust grain model includes optical properties of graphite and polycyclic aromatic hydrocarbons (PAHs) for carbonaceous dust and olivine optical properties for astronomical silicates \citep{Draine84,Li01,Weingartner01,Clayton03,Siebenmorgen14,Zhukovska18}.

After hydrogen, helium, carbon and oxygen, iron is the fifth most abundant element by mass \citep{Asplund09} and the third largest contributor to the interstellar dust mass after carbon and oxygen \citep{Jenkins09,Draine11}. Nanoparticles discovered in meteorites indicate the interstellar dust grain-size distribution extends down to nanometers \citep{Mautner06}. However, when modeling nanoparticles it is challenging to take into account the many effects of the environment on nanoparticles' erosion, mantling, and coagulation. It is also difficult to account for the return effect nanoparticles have on the environment, such as photoelectric heating of the gas, formation of molecules and small radicals by surface reactions \citep{Jones16}. Interstellar iron nanoparticles have previously been accounted for within silicates, particularly in olivines and pyroxenes \citep{Westphal14,Kohler14,Hilchenbach16}. \citet{Draine13} and \citet{Hensley17} discussed metallic iron and iron oxides as free-flying nanoparticles and inclusions within larger grains as contributing to interstellar dust. Metallic iron's high abundance and contribution to far-ultraviolet extinction makes it a potential candidate for contributing to interstellar extinction and reddening \citep{Kohler14}.

Dust production and destruction in the distant universe is still under debate based on the timeline required for dust production \citep{Todini01,Morgan03,Hirashita05,Schneider12,Valiante09,Zafar18b}. At higher redshifts, individual extinction curves of $\gamma$-ray bursts (GRBs) have been estimated for several cases usually finding an extinction curve steeper than the SMC \citep{Zafar18b}. From the depletion and extinction analysis of a sample of GRBs, \citet{Zafar19b} suggested that high silicon and iron depletions indicate a variation in dust grain-sizes which can alter the shape of the individual extinction curves.

In this work, we aim to find dust grain-size distributions that include iron nanoparticles in addition to graphite, PAHs, and astronomical silicates by fitting a model to an extinction curve exhibiting a 2175\,\AA\ bump in a distant GRB afterglow environment. We use the recently discovered GRB\,180325A which is the first GRB in a decade that contains a 2175\,\AA\ bump in its extinction curve reported by \cite{Zafar18}. They performed a spectroscopic spectral energy distribution (SED) fitting of GRB\,180325A at four different epochs. This paper is organised as follows: in \S2 we provide observational extinction curve data of GRB\,180325A. We discuss the method for fitting the observed extinction curve in \S3. We present the results in \S4 and give a discussion of the contribution of different grain types and dust destruction in \S5. The conclusions are provided in \S6.

\section{Dust in GRB~180325A}
A characteristic feature in the Milky Way is the 2175\,\AA\ bump that is ubiquitously detected in the Galaxy but is rarely observed in the distant universe. The 2175\,\AA\ bump has been spectroscopically confirmed in five GRB environments \citep{Zafar12,Zafar18}. The fifth recent unambiguous spectroscopic detection of a 2175\,\AA\ bump is reported towards the line of sight of a GRB afterglow, GRB\,180325A at $z=2.2486$ \citep{Zafar18}\footnote{Note that recently, \cite{deUgartePostigo21} reported the detection of an unambiguous 2175 {\AA} bump in a \emph{foreground absorber} along the line of sight toward GRB 210619B.}. The bump has been spectroscopically observed at four different epochs up to $\sim$3 hours after the burst trigger. All four SEDs are well-described by a 2175\,\AA\ bump with $R_V \approx 4.4$, $A_V\approx 1.5$\,mag, and log\,N({H\,{\sc i}}/cm$^{-2}$) $= 22.30$. Later for a sample of GRBs (including GRB\,180325A), \citet{Heintz19} found that the rarely detected neutral atomic-carbon in GRBs is correlated with $R_V$ and $A_V$. These three quantities are connected with the strength of the 2175\,\AA\ bump, suggesting that the bump is produced by the carbon-rich dust in the molecular cloud \citep{Heintz19}. 

In this work, we use the third epoch extinction curve of GRB\,180325A at $\Delta t=1.63$\,hrs obtained from the spectroscopic X-ray to near-infrared SED fit. We chose the extinction curve from this epoch because of highest signal-to-noise ratio, broad wavelength coverage, and the best fit of the four epochs in \citet{Zafar18}. For this epoch X-ray data is obtained from the \emph{Swift} satellite and the broad wavelength coverage ultraviolet to near-infrared spectrum is acquired with the X-shooter instrument mounted at the Very Large Telescope, European Southern Observatory, Chile. For the $\Delta t=1.63$\,hrs epoch, \citet{Zafar18} reported that the observed SED can be described by $R_V=4.58^{+0.37}_{-0.39}$ and restframe $V$-band extinction of $A_V=1.58^{+0.10}_{-0.12}$\,mag. The neutral hydrogen column density derived from the Ly$\alpha$ absorption-line is log\,N({H\,{\sc i}}/cm$^{-2}$) $= 22.30$ \citep{Zafar18}. We use these three values to fit an interstellar dust grain model to the observed extinction curve of GRB\,180325A.

\section{Method}
This study uses a similar methodology to \citet[][hereafter~WD2001]{Weingartner01} to fit carbonaceous (graphite and PAHs) and silicates dust grain-size distributions containing a substantial contribution from very small grains. 
The simplest dust models have been demonstrated to reproduce observed extinction curves for a wide range of local universe environments (\citealt{Draine84,Zubko98,Li01}; WD2001). Depletion patterns suggest that metallic iron and iron oxides are an important component of dust \citep{Sofia94,Jenkins09,Draine13}. Our dust model incorporates free-flying metallic iron nanoparticles from \citet{Draine13} for the first time to explain the dust in a GRB environment. We aim to find whether a mix of carbonaceous, silicates and iron nanoparticles are required to explain the observed extinction curve of the afterglow of GRB\,180325A.

\subsection{Carbonaceous grain-size distribution}
For graphite, we use dust grain-size distribution equations defined in WD2001 and \citet{Li01}: 
\begin{equation}
    \frac{1}{n_{\text{H}}}\frac{dn_g}{da} =  \frac{C_g}{a} \left( \frac{a}{a_{t,g}}\right)^{\alpha_g} F\left(a;\beta_g, a_{t,g} \right ) \times G\left(a;a_{t,g},a_{c,g} \right)
\end{equation}

where:
\begin{equation}
    F \left (a;\beta , a_t \right) \equiv \Bigg \{ 
    \begin{array}{lll}
        1 + \beta a/a_t & & \beta\geq0 \\
        \left(1 - \beta a/a_t\right)^{-1} & & \beta<0 \\
\end{array}  
\end{equation}
\begin{equation}
  G \left (a;a_t , a_c \right) \equiv \Bigg \{ 
\begin{array}{ll}
    1 & 3.5\,\mbox{\normalfont\AA} < a < a_t \\
    {\text{exp}}\{-\left[\left(a-a_t\right)/a_c\right]^3\} & a > a_t\\
\end{array}  
\end{equation}

Equation 1 provides five adjustable parameters ($C_g$, $a_{t,g}$, $\alpha_g$, $\beta_g$, and $a_{c,g}$) for the graphite dust grain-size distribution. Smaller carbonaceous dust grains in the form of PAHs have a log-normal size distribution:
\begin{equation}
        \frac{1}{n_{\text{H}}}\frac{dn_{\text{PAH}}}{da} = \sum_{i=1}^{2} \frac{B_i}{a} {\text{exp}} \Bigg \{ -\frac{1}{2} \Bigg [ \frac{\text{ln} (a/a_{0,i})}{\sigma} \Bigg]^2 \Bigg \}
\end{equation}

for $a>3.5$\,\AA\ where $B_i$ is defined as:
\begin{equation}
    B_i = \frac{3}{\left(2\pi\right)^{3/2}} \frac{\text{exp}(-4.5\sigma^2)}{a_{0,i}^3\rho\sigma} \frac{b_{\text{C,\it{i}}}\, m_{\text{C}}}{1+\text{erf}[3\sigma/\sqrt{2} + \ln (a_{0,i}/a_{\text{min}})/\sigma\sqrt{2}]}
\end{equation}
where $a$ is the dust grain radius in cm, $m_{\text{C}} = 1.94 \times 10^{-23}$\,g is the mass of a carbon atom, $\rho = 2.266$\,g cm$^{-3}$ is the density of graphite, $b_{\text{C,1}} = 0.75b_C$, $b_{\text{C,2}} = 0.25b_C$ where $b_C$ is the total carbon abundance relative to hydrogen. We adopt $a_{\text{min}} = a_{0,1} = 3.5$\,\AA, $a_{0,2} = 30$\,\AA\ and  $\sigma = 0.4$, as used in WD2001.

The parameter $b_C$ is adjustable. WD2001 modelled several grain-size distributions of Milky Way extinction curves with different values of $b_C$ ranging from $0.0 - 6.0 \times 10^{-5}$  at integer intervals. However for LMC\,2 with a smaller bump, WD2001 used $b_C$ values of (0.0, 0.5, 1.0) $\times10^{-5}$. As GRB\,180325A's extinction curve is quite steep in the UV regime, similar to the SMC's extinction curve, we first fitted for values of $b_C=$(0.0, 0.5, 1.0, 2.0)$\times10^{-5}$. We find that $b_C=0.5\times10^{-5}$ provides the better fit among all scenarios, therefore, we fit for various values of $b_C$ between $0 - 0.5\times 10^{-5}$  at intervals of $0.1\times 10^{-5}$. 

\subsection{Silicates grain-size distribution}
For silicates dust, the grain-size distribution is given as:
\begin{equation}
  \frac{1}{n_{\text{H}}}\frac{dn_{\text{sil}}}{da} =  \frac{C_s}{a} \left( \frac{a}{a_{t,s}}\right)^{\alpha_s} F\left(a;\beta_s, a_{t,s} \right ) \times G\left(a;a_{t,s},a_{c,s} \right)  
\end{equation}
The terms $F\left(a;\beta_s, a_{t,s} \right )$ and $G\left(a;a_{t,s},a_{c,s}\right)$ are similar as in equations 2 and 3. Equation 6 provides four adjustable parameters ($C_s$, $a_{t,s}$, $\alpha_s$, and $\beta_s$) for the silicates size distribution. The parameters $a_{c,s}$ is fixed to $0.1$\,$\mu$m as in WD2001.

\subsection{Iron nanopartices size distribution}
More than 95\% of iron is depleted from the gas phase \citep{Jenkins09} and this depletion can be explained by the accretion of iron onto dust grains \citep{Dwek16}. A large fraction of iron is locked in silicates \citep{Westphal14} and can be in the form of metallic iron \citep{Schalen65}, in silicates lattices \citep{Ossenkopf92}, pure iron and iron-sulphates \citep{Jones13}, as iron oxides \citep{Draine13}, or in the form of free-flying iron nanoparticles \citep{Hoang16,Hensley17,Gioannini17,Bilalbegovic17}. In this work we attempted for the first time to take into account the contributions of iron nanoparticles and iron oxides (Magnetite Fe$_3$O$_4$ and Maghemite $\gamma$-Fe$_2$O$_3$) and study whether their inclusion can explain the far-UV rise of the extinction curve. We used the optical properties provided in \citet{Draine13} and \cite{Hensley17} (hereafter~HD2017) for metallic iron, Magnetite Fe$_3$O$_4$ and Maghemite $\gamma$-Fe$_2$O$_3$.

\begin{figure}
\begin{center}
{\includegraphics[width=\columnwidth,bb=45 85 855 588]{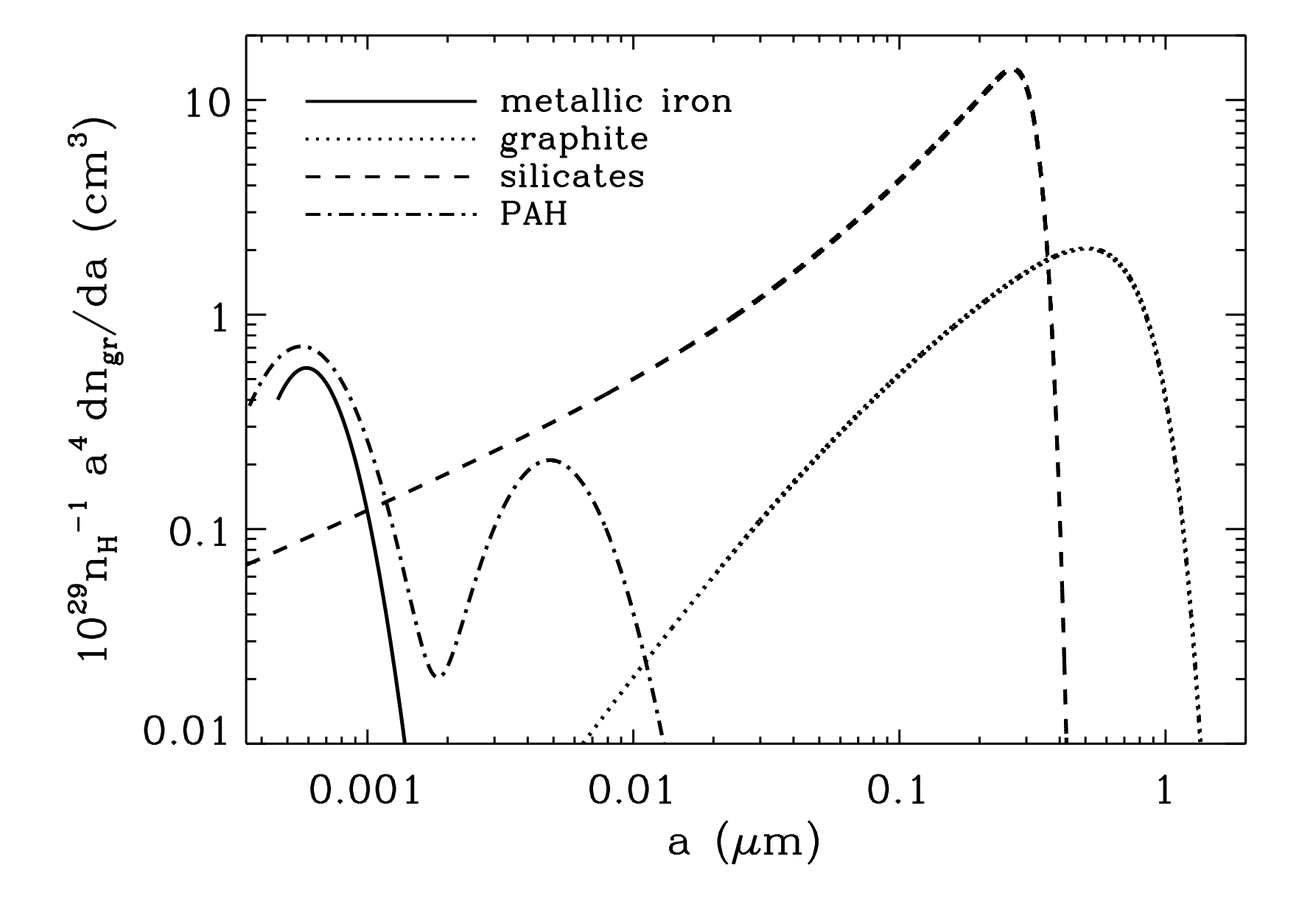}}
\caption{An illustrative example of the SPGI dust grain-size distributions for the GRB\,180325A environment to present grain-size ranges for each component. The best-fit model favoured distributions given in Table \ref{table:1} are shown where the solid curve corresponds to metallic iron ($Y_{\text{Fe}}=0.03$); the dotted to graphite; the dashed to silicates, and the dot-dashed to hydrocarbon molecules ($b_C=0.4$\,$\times10^{-5}$).}
\label{dnda}
\end{center}
\end{figure}

The assumed log-normal size distribution for free-flying iron nanoparticles is given as:
\begin{equation}
    \frac{1}{n_{\text{H}}}\frac{dn_{\text{Fe}}}{da} = \frac{A}{a}\text{exp} \Bigg \{ -\frac{1}{2} \Bigg [\frac{\text{ln}(a/a_0)}{\Tilde{\sigma}} \Bigg ]^2 \Bigg \}
\end{equation}
where the normalisation constant is defined as:
\begin{equation}
    A = \frac{3}{\left(2\pi\right)^{3/2}} \frac{\text{exp}(-4.5\Tilde{\sigma}^2)}{a_{0}^3\rho\Tilde{\sigma}} \times \frac{m_{\text{Fe}}b_{\text{Fe}}}{1+\text{erf}[3\Tilde{\sigma}/\sqrt{2} + \ln (a_{0}/a_{\text{min}})/\Tilde{\sigma}\sqrt{2}]}
\end{equation}
where $a$ is the dust grain radius in cm, $\rho = 7.87$\,g cm$^{-3}$ is the density of pure iron, $m_{\text{Fe}} = 9.27 \times 10^{-23}$\,g is the mass of an iron atom, $b_{\text{Fe}} = 41 Y_{\text{Fe}} \times 10^{-6}$  is the abundance of iron atoms per hydrogen where $Y_{\text{Fe}}$ is the fraction of iron atoms that are in free-flying nanoparticles (see HD2017).

We use $\Tilde{\sigma} = 0.3$ with $a_0 = 4.5$\,\AA\ in equation 7 from the illustrative example in HD2017. Note that \citet{Hoang16} used equation 7 with $\Tilde{\sigma} = 0.3$ with $a_0 = 4.0$\,\AA\ and $\Tilde{\sigma} = 0.6$ with $a_0 = 40$\,\AA . The grain radii sizes available from \citet{Draine13} for metallic iron ranged from 10--10,000\,\AA, therefore, we chose to adopt $\Tilde{\sigma} = 0.3$ and $a_0 = 4.5$\,\AA. The proportion of iron existing as free-flying nanoparticles is unknown (HD2017) so we fitted for a range of values of $Y_{\text{Fe}}$ from $0.0 - 0.1$ in increments of $0.01$.

We considered including iron oxides Magnetite (with $\rho = 5.20$\,g cm$^{-3}$ and $m_{\text{Fe$_3$O$_4$}} = 3.88 \times 10^{-22}$\,g) and Maghemite (with $\rho = 4.899$\,g cm$^{-3}$ and $m_{\text{$\gamma$-Fe$_2$O$_3$}} = 2.65 \times 10^{-22}$\,g) in our model, but the optical properties in \citet{Draine13} provided for the size distributions of large grain radii make no significant contributions to the modelled extinction curve. Therefore, we excluded iron oxide components from our model as not enough grain sizes range is available to constrain their contribution.

\subsection{Model extinction curve}
The total extinction at wavelength $\lambda$ is given as:
\begin{equation}
    A(\lambda) = N_H 2.5 \pi \log(\text{e}) \int \frac{1}{n_{\text{H}}} \frac{dn_{\text{grain}}}{da} Q_{\text{ext}}(a,\lambda) a^2 da
\end{equation}
where $N_H=10^{22.3}\text{cm}^{-2}$ is the hydrogen column density, obtained from \citet{Zafar18}, $\frac{1}{n_{\text{H}}}\frac{dn}{da}$ is the grain-size distribution, as defined in equations 1 to 8 for different grain compositions. $Q_{\text{ext}}$ is the extinction efficiency coefficient for each grain radius and incident wavelength. The values for $Q_{\text{ext}}$ have been calculated using Mie theory.
We used the graphite, PAHs, and `smoothed astronomical silicate' optical properties calculated by WD2001\footnote{ https://www.astro.princeton.edu/$\sim$draine/}. We assume that PAHs are $50\%$ neutral and $50\%$ ionised. The graphite optical properties use the approximation where the dielectric tensor for the electric field perpendicular  and parallel to the c-axis make contributions of 1/3 and 2/3 to $Q_{\text{ext}}$  respectively.

We aim to find the best-fit model for GRB\,180325A extinction data for given values of $b_C$ and $Y_{\text{Fe}}$ and fitting the graphite and silicates grain-size distribution parameters. We used the Levenberg-Marquardt method for nonlinear least squares curve-fitting to fit the GRB\,180325A extinction curve from 0.001 to 8\,$\mu$m$^{-1}$. The extinction curve of GRB\,180325A was interpolated to match the optical properties data from WD2001 for fitting purposes resulting in 785 data points. The fit was performed within \texttt{IDL} using the \texttt{MPFITFUN} module. The infrared region of the extinction curve was fitted by interpolating fixed spline points using the \citet{fm07} model. This resulted in very small errors for that part of the extinction curve. We provided the same weighting treatment to errors in the infrared region as in WD2001 to allow the fitting algorithm to find the best solution. The errors in the infrared region are quite small, therefore, we set the weights $\sigma_i^{-1} = 1/10$ for $\lambda^{-1} < 1.1\mu\text{m}^{-1}$ due to the uncertainty of the actual IR extinction (see WD2001). 

We fit four grain models with different combinations of dust materials to determine the significance of each type of dust when fitting the GRB extinction curve. We first fit a dust grain model with silicates, PAHs, graphite and iron nanoparticles (SPGI model) and silicates, PAHs and graphite (SPG model). Both of these models have 9 free parameters (five for graphite, four for silicates) as mentioned above for each grain-size distribution. These models have 776 degrees of freedom.

We also fit models without graphite to determine if the 2175\,{\AA} bump was best fit by PAHs alone. We fit silicates, PAHs and iron nanoparticles (SPI model) and silicates and PAHs (SP model). Both of these models have 4 free parameters and 781 degrees of freedom. The SP model was to test the significance of an iron contribution.

We fit the the extinction curve of GRB\,180325A with the SPGI model for $b_C= 0.0$ to $1.0\,\times10^{-5}$ with a step-size of $0.1\times10^{-5}$ and $Y_{\text{Fe}}= 0.0$ to $0.1$ with a step-size of $0.01$. This made 110 cases of varying $b_C$ and $Y_{\text{Fe}}$ and for each set we fitted the extinction curve of GRB\,180325A and estimated the best-fit parameters. We later fit the extinction curve of GRB\,180325A with the SPG (11 cases), SPI (100 cases) and SP (11 cases) models. We fit these four different permutations for 232 sets of $b_C$ (with $b_C=0.0 -1.0 \times 10^{-5}$ with a step-size of $0.1\times10^{-5}$) and $Y_{\text{Fe}}$ (with $Y_{\text{Fe}}=0.0-0.1$ with a step-size of $0.01$).

\section{Results}
All four model (namely, SP, SPI, SPG and SPGI) parameter permutations provided multiple fits for the GRB\,180325A extinction curve each with p-values < 0.05 and $1 < \chi^2_\nu < 2$ for different ranges of $b_C$ and $Y_{\rm Fe}$, as provided in Table \ref{table:1}. The carbon abundance, $b_C$, is significantly degenerated for each permutation.
In addition, we find that SPI and SPGI models present an important degeneracy regarding the iron abundance, $Y_{\rm Fe}$.
The full list containing the 232 fits across all four models is provided in Table \ref{table:2}.

The simplest SP model provides best fits for $b_C=0.3 - 0.7 \times 10^{-5}$, whereas for $b_C=0 - 0.2, 0.8 - 1.0 \times 10^{-5}$ model indicates poor fits with $\chi^2_\nu > 2$. The SPI model provides best fits for $b_C=0.3 - 0.6 \times 10^{-5}$, each with multiple Y$_\text{Fe}$ values (see Table \ref{table:1}). The models with $b_C < 0.2 \times 10^{-5}$ and $b_C > 0.7 \times 10^{-5}$ resulted poor fits with $\chi^2_\nu > 2$.

The SPG model fits well with $b_C=0.0, 0.4 - 0.7 \times 10^{-5}$. Models with $b_C=0.1 - 0.3, 0.8 - 1.0 \times 10^{-5}$ resulted in $\chi^2_\nu > 2$, indicating a poor fit. The SPGI model had 20 permutations of $b_C$ and Y$_\text{Fe}$ which are fit best with p-value $< 0.05$ and $1 < \chi^2_\nu < 2$ All other SPGI models were poor fits with $\chi^2_\nu >2$.

\begin{figure}
\begin{center}
{\includegraphics[width=\columnwidth,bb=35 85 915 620]{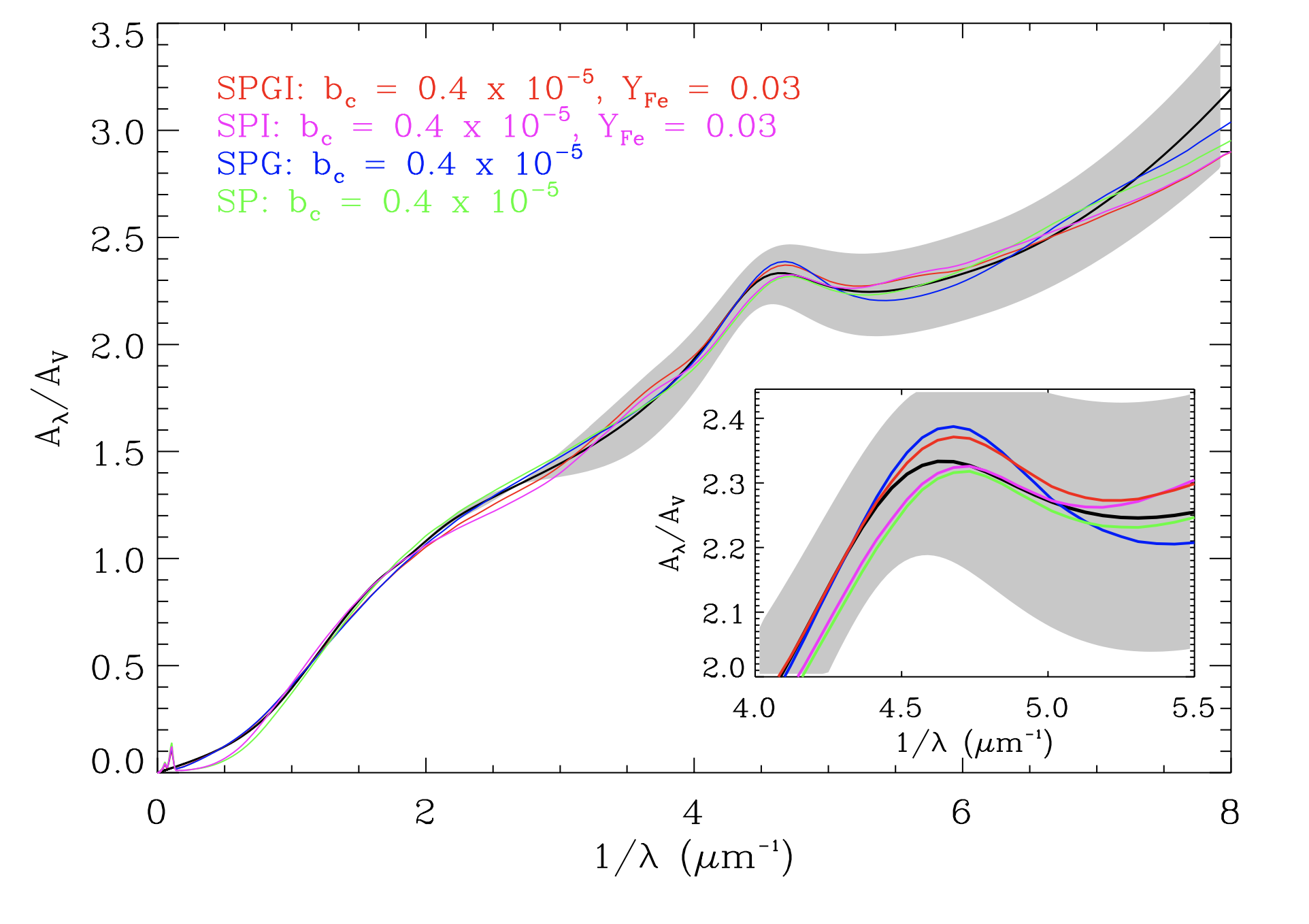}}
\caption{Modelled extinction curves overlaid on observed extinction curve (solid black line). The gray-shaded area represents the 1$\sigma$ uncertainty on the observed extinction curve. The modelled extinction curves for the best-fit SPGI ($b_C, Y_{\rm Fe}=0.4\times10^{-5}, 0.03$), SPI ($b_C, Y_{\rm Fe}=0.4\times10^{-5}, 0.03$), SPG ($b_C=0.4$\,$\times10^{-5}$) and SP ($b_C=0.4$\,$\times10^{-5}$) models are shown in red, magenta, blue, and green colours, respectively.}
\label{multi}
\end{center}
\end{figure}

As an example, the contributions from very small hydrocarbon molecules, metallic iron nanoparticles, silicates and graphite grain distributions at a wide range of grain-sizes for the SPGI with $b_C=0.4\times10^5$ and $Y_{\text{Fe}}=0.03$ are shown in Fig.\,\ref{dnda}. Illustrative examples of best-fits for all four models are shown in Fig. \ref{multi}. These models included are the following: SPGI ($b_C, Y_{\rm Fe}=0.4\times10^{-5}, 0.03$), SPI ($b_C, Y_{\rm Fe}=0.4\times10^{-5}, 0.03$), SPG ($b_C=0.4$\,$\times10^{-5}$) and SP ($b_C=0.4$\,$\times10^{-5}$). It is worth noting that none of the proposed models are able to reproduce the steep trend of the observed extinction curve at the far-UV regime ($\sim 1400$ \AA), although the uncertainties in the data preclude us from any robust conclusion. Further examples of fits from all four model permutations are provided in the appendix (SP: Figure \ref{fig:multidustsp}, SPG: Figure \ref{fig:multidustspg},  SPI: Figure \ref{fig:multidustspi}, SPGI: Figure \ref{fig:multidustspgi}).

\begin{table*}
\caption{Fitted grain-size distribution parameter values for GRB\,180325A. For each model, $\chi^2_\nu$ and p-values are provided. The number of degrees of freedom for each model is included along with the adopted name.}
\label{table:1} 
\centering     
\setlength{\tabcolsep}{6pt}
\begin{tabular}{c c c c c c c c c c c c c }  
\hline\hline                        
$b_C$ & $Y_{\text{Fe}}$ & $\alpha_g$ & $\beta_g$ & $a_{t,g}$ & $a_{c,g}$ & $C_g$ & $\alpha_s$ & $\beta_s$ & $a_{t,s}$ & $C_s$ & $\chi^2_\nu$ & p-value\\
$10^{-5}$ &  &  &  & $\mu$m & $\mu$m & $10^{-16}$ &  &  & $\mu$m & $10^{-16}$ &  &  \\
\hline
\textbf{SP} (781)& \\
0.3 & $\cdots$ & $\cdots$ & $\cdots$ & $\cdots$ & $\cdots$ & $\cdots$ & -2.89 & 22.92 & 0.23 & 5.35 & 1.99 & 0.00 \\
0.4 & $\cdots$ & $\cdots$ & $\cdots$ & $\cdots$ & $\cdots$ & $\cdots$ & -2.89 & 19.49 & 0.25 & 4.83 & 1.48 & 0.00 \\
0.6 & $\cdots$ & $\cdots$ & $\cdots$ & $\cdots$ & $\cdots$ & $\cdots$ & -2.74 & 6.85 & 0.30 & 7.79 & 1.52 & 0.00 \\
0.7 & $\cdots$ & $\cdots$ & $\cdots$ & $\cdots$ & $\cdots$ & $\cdots$ & -2.77 & 7.24 & 0.34 & 4.93 & 1.90 & 0.00 \\
\hline
\textbf{SPI} (781)& \\
0.3 & 0.01 & $\cdots$ & $\cdots$ & $\cdots$ & $\cdots$ & $\cdots$ & -2.83 & 20.95 & 0.24 & 5.19 & 1.75 & 0.00 \\
0.3 & 0.02 & $\cdots$ & $\cdots$ & $\cdots$ & $\cdots$ & $\cdots$ & -2.70 & 13.74 & 0.24 & 7.44 & 1.72 & 0.00 \\
0.3 & 0.03 & $\cdots$ & $\cdots$ & $\cdots$ & $\cdots$ & $\cdots$ & -2.53 & 8.07 & 0.24 & 12.07 & 1.78 & 0.00 \\
0.3 & 0.04 & $\cdots$ & $\cdots$ & $\cdots$ & $\cdots$ & $\cdots$ & -2.36 & 6.07 & 0.24 & 15.66 & 1.92 & 0.00 \\
0.4 & 0.01 & $\cdots$ & $\cdots$ & $\cdots$ & $\cdots$ & $\cdots$ & -2.77 & 12.69 & 0.26 & 6.70 & 1.28 & 0.00 \\
0.4 & 0.02 & $\cdots$ & $\cdots$ & $\cdots$ & $\cdots$ & $\cdots$ & -2.63 & 7.98 & 0.27 & 9.58 & 1.26 & 0.00 \\
0.4 & 0.03 & $\cdots$ & $\cdots$ & $\cdots$ & $\cdots$ & $\cdots$ & -2.52 & 6.92 & 0.27 & 9.96 & 1.41 & 0.00 \\
0.4 & 0.04 & $\cdots$ & $\cdots$ & $\cdots$ & $\cdots$ & $\cdots$ & -2.40 & 6.27 & 0.28 & 10.65 & 1.64 & 0.00 \\
0.4 & 0.05 & $\cdots$ & $\cdots$ & $\cdots$ & $\cdots$ & $\cdots$ & -2.17 & 5.07 & 0.26 & 14.93 & 1.93 & 0.00 \\
0.4 & 0.07 & $\cdots$ & $\cdots$ & $\cdots$ & $\cdots$ & $\cdots$ & -0.87 & 2.44 & 0.19 & 54.50 & 1.97 & 0.00 \\
0.4 & 0.08 & $\cdots$ & $\cdots$ & $\cdots$ & $\cdots$ & $\cdots$ & 0.61 & 6.70 & 0.15 & 23.76 & 1.94 & 0.00 \\
0.5 & 0.02 & $\cdots$ & $\cdots$ & $\cdots$ & $\cdots$ & $\cdots$ & -2.64 & 7.54 & 0.31 & 6.84 & 1.20 & 0.00 \\
0.5 & 0.03 & $\cdots$ & $\cdots$ & $\cdots$ & $\cdots$ & $\cdots$ & -2.57 & 7.55 & 0.32 & 5.89 & 1.45 & 0.00 \\
0.5 & 0.04 & $\cdots$ & $\cdots$ & $\cdots$ & $\cdots$ & $\cdots$ & -2.45 & 7.11 & 0.33 & 6.05 & 1.80 & 0.00 \\
0.5 & 0.08 & $\cdots$ & $\cdots$ & $\cdots$ & $\cdots$ & $\cdots$ & 2.79 & 0.85 & 0.13 & 50.35 & 1.88 & 0.00 \\
0.6 & 0.01 & $\cdots$ & $\cdots$ & $\cdots$ & $\cdots$ & $\cdots$ & -2.72 & 7.58 & 0.33 & 5.52 & 1.39 & 0.00 \\
0.6 & 0.02 & $\cdots$ & $\cdots$ & $\cdots$ & $\cdots$ & $\cdots$ & -2.67 & 7.87 & 0.36 & 4.33 & 1.55 & 0.00 \\
0.6 & 0.03 & $\cdots$ & $\cdots$ & $\cdots$ & $\cdots$ & $\cdots$ & -2.58 & 7.34 & 0.37 & 4.22 & 1.95 & 0.00 \\
\hline
\textbf{SPG} (776)& \\
0.0 & $\cdots$ & -2.21 & -0.87 & 0.10 & 0.78 & 256.52 & -3.27 & 130.29 & 0.24 & 0.52 & 1.18 & 0.00 \\
0.4 & $\cdots$ & -1.68 & -1.03 & 0.08 & 0.66 & 271.71 & -2.53 & 1.51 & 0.23 & 36.86 & 1.10 & 0.03 \\
0.5 & $\cdots$ & -1.39 & -0.82 & 0.07 & 0.62 & 260.09 & -2.61 & 2.51 & 0.23 & 27.33 & 1.17 & 0.00 \\
0.6 & $\cdots$ & -1.00 & -0.58 & 0.05 & 0.60 & 171.87 & -2.67 & 3.84 & 0.24 & 18.82 & 1.38 & 0.00 \\
0.7 & $\cdots$ & -0.66 & -0.35 & 0.03 & 0.60 & 127.52 & -2.69 & 4.45 & 0.25 & 14.75 & 1.66 & 0.00 \\
\hline
\textbf{SPGI} (776)& \\
0.0 & 0.02 & -2.17 & -1.78 & 0.18 & 0.84 & 72.88 & -2.63 & 3.28 & 0.22 & 20.60 & 1.09 & 0.04 \\
0.0 & 0.03 & -2.11 & -2.12 & 0.18 & 0.85 & 76.04 & -2.56 & 3.84 & 0.22 & 18.87 & 1.11 & 0.02 \\
0.0 & 0.04 & -2.06 & -2.39 & 0.19 & 0.89 & 73.79 & -2.43 & 3.86 & 0.22 & 20.19 & 1.50 & 0.00 \\
0.0 & 0.05 & -2.02 & -3.20 & 0.23 & 0.93 & 50.09 & -2.23 & 3.61 & 0.21 & 23.74 & 1.49 & 0.00 \\
0.1 & 0.04 & -1.93 & -2.27 & 0.16 & 0.85 & 94.54 & -2.40 & 3.79 & 0.22 & 21.31 & 1.28 & 0.00 \\
0.1 & 0.05 & -1.89 & -2.89 & 0.20 & 0.88 & 65.41 & -2.18 & 3.49 & 0.21 & 25.57 & 1.43 & 0.00 \\
0.2 & 0.04 & -1.77 & -2.34 & 0.17 & 0.78 & 86.11 & -2.36 & 3.79 & 0.22 & 22.36 & 1.18 & 0.00 \\
0.2 & 0.05 & -1.71 & -2.62 & 0.18 & 0.82 & 76.05 & -2.14 & 3.52 & 0.21 & 26.80 & 1.28 & 0.00 \\
0.3 & 0.03 & -1.62 & -1.21 & 0.09 & 0.74 & 212.61 & -2.49 & 4.02 & 0.22 & 20.41 & 1.12 & 0.01 \\
0.3 & 0.04 & -1.53 & -2.81 & 0.20 & 0.67 & 53.61 & -2.32 & 3.86 & 0.22 & 23.15 & 1.30 & 0.00 \\
0.3 & 0.05 & -1.43 & -2.29 & 0.16 & 0.74 & 69.41 & -2.08 & 3.46 & 0.21 & 28.79 & 1.56 & 0.00 \\
0.4 & 0.03 & -1.29 & -0.94 & 0.08 & 0.69 & 168.06 & -2.45 & 4.13 & 0.22 & 20.49 & 1.24 & 0.00 \\
0.4 & 0.04 & -1.11 & -1.40 & 0.11 & 0.68 & 78.70 & -2.28 & 3.88 & 0.22 & 23.62 & 1.62 & 0.00 \\
0.4 & 0.05 & -0.96 & -2.51 & 0.20 & 0.61 & 34.54 & -2.01 & 3.50 & 0.21 & 29.44 & 1.82 & 0.00 \\
0.5 & 0.02 & -1.05 & -0.68 & 0.06 & 0.65 & 156.76 & -2.55 & 4.34 & 0.23 & 18.07 & 1.13 & 0.01 \\
0.5 & 0.03 & -0.85 & -0.86 & 0.07 & 0.66 & 80.44 & -2.42 & 4.20 & 0.23 & 19.41 & 1.51 & 0.00 \\
0.5 & 0.04 & -0.59 & -1.91 & 0.15 & 0.61 & 35.64 & -2.23 & 3.98 & 0.23 & 22.24 & 1.81 & 0.00 \\
0.6 & 0.01 & -0.85 & -0.59 & 0.05 & 0.62 & 114.69 & -2.63 & 4.46 & 0.24 & 16.11 & 1.36 & 0.00 \\
0.6 & 0.02 & -0.68 & -0.65 & 0.06 & 0.64 & 71.76 & -2.53 & 4.49 & 0.24 & 16.04 & 1.54 & 0.00 \\
0.7 & 0.01 & -0.56 & -1.00 & 0.09 & 0.60 & 47.20 & -2.63 & 4.70 & 0.25 & 13.64 & 1.95 & 0.00 \\

\hline 
\end{tabular}
\end{table*}

The more complex models, including graphite and iron nanoparticles, do not significantly improve the fit of the extinction curve, therefore, we favour the SP model as it is the simplest model. The results of all these fits from different model permutations indicate that both the UV regime and bump can be explained by either PAHs or graphite and iron nanoparticles do not contribute significantly to explain the UV rise for GRB\,180325A. 

For GRB\,180325A, the carbon abundance range of the best fit models ($0.0 \leq b_C \leq 0.7) \times 10^{-5}$ is smaller compared to the previous models of the Milky Way sightlines (e.g., \citealt{Draine84}; WD2001; \citealt{Muthumariappan10}). Note that a smaller bump and flatter extinction curve compared to the Milky Way is reported for GRB\,180325A \citep{Zafar18}.

\section{Discussions}
\subsection{Iron nanoparticle contribution}

Our results indicate contribution of iron nanoparticles is not significant, therefore, iron contribution is not necessary to explain the observed extinction curve of GRB\,180325A. At smaller wavelengths, from $7$\,$\mu\text{m}^{-1}$  the model and observed $A(\lambda)$ diverge. At this point the steepness of the modelled silicates extinction starts to decrease while the observed $A(\lambda)$ curve increases in steepness. As nanoparticles strongly contribute to the UV extinction, inclusion of optical properties data from silicates nanoparticles and for more grain-sizes for iron oxide nanoparticles might be capable of improving the dust grain models and better fit the steep UV rise in GRB\,180325A.

Iron nanoparticles are an important component of interstellar dust. Metallic iron has been found in interplanetary dust \citep{Bradley94}. Inclusions of metallic iron consistent with iron nanoparticles have been identified in the interstellar grains collected in the Solar System \citep{Westphal14}. Iron nanoparticles may contribute to interstellar heating (WD2001). The anomalous microwave emission in the 10--60\,GHz frequency range is attributed to iron nanoparticles through thermal magnetic dipole emission \citep{Draine13} or rotational emission \citep{Hoang16}. HD2017 found that iron nanoparticles make a minor contribution to the total photoelectric heating of ambient interstellar gas in that region provided that $Y_{\text{Fe}} \gtrapprox 10\%$. Due to negligible iron contribution, we cannot conclude whether iron nanoparticles are contributing to the photoelectric heating of the interstellar gas in that region. Future dust grain models of different extinction curves may benefit from including iron nanoparticle grains of radii smaller than $1\times10^{-3} \mu$m. 

\subsection{Graphite and PAHs contribution}

We fit the GRB\,180325A extinction curve with and without graphite. The data is best fit by a range of $b_C$ values, with ($0.0 \leq b_C \leq 0.7)\times10^{-5}$. The models with and without graphite show a degeneracy in the carbon abundance ($b_C$) and similar carbon degeneracy for the Milky Way sightlines is reported by WD2001. In all four models, the $\chi^2_{\nu}$ does not improve when graphite is included which implies that including graphite does not significantly improve the fit.

The 2175\,\AA\ bump feature can be produced by both graphite and PAH. WD2001 found that the 2175\,\AA\ bump is contributed to almost completely by PAHs in the MW. This could be the case in other galaxies. In this work however, we did not find a great improvement when we modelled GRB 180325A’s extinction curve with graphite. Ultimately, as the more complex models that include graphite do not significantly improve the fit of the GRB 180325A extinction curve, we favour the simpler SP model. Therefore, the 2175 {\AA} bump is best fit by PAHs. \citet{Ma20} proposed that hydrogenated T-carbon (HTC) molecule mixtures may cause the 2175 \,\AA\ bump. Incorporating other forms of carbon, such as HTCs, in dust grain distribution models will help towards further understanding of the origin of the 2175\,\AA\ bump. 

\subsection{Dust destruction}

\begin{figure}
\begin{center}
{\includegraphics[width=\columnwidth]{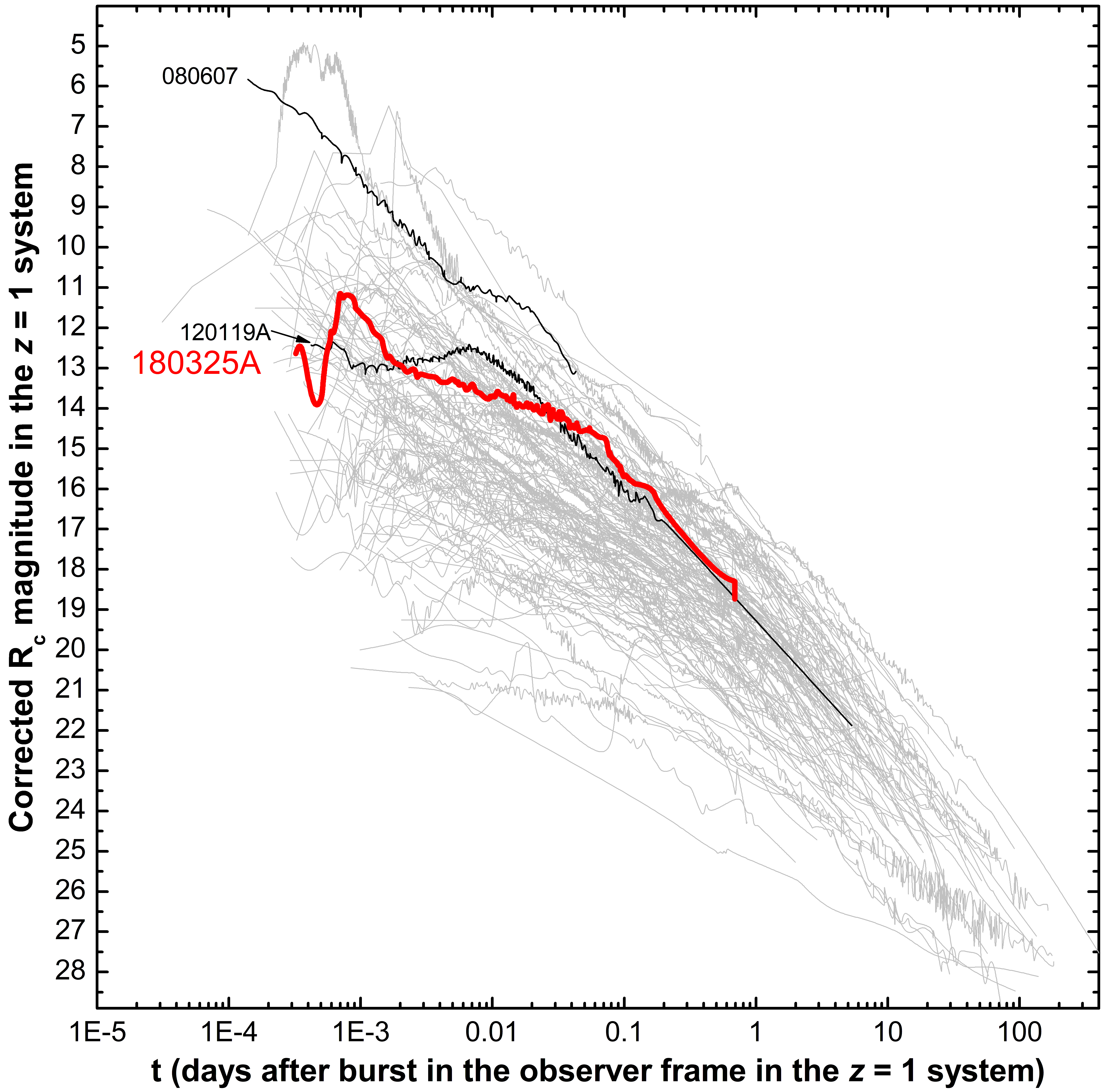}}
\caption{The afterglow light curve of GRB\,180325A (red) after correction for all extinction and a shift to $z=1$, in comparison to a large sample of GRB afterglows (\citealt{Kann06,Kann10,Kann11}, Kann et al., in prep.). We also highlight the ultraluminous GRB\,080607, one of the most luminous early afterglows ever discovered, which showed high extinction but no evidence for dust destruction, as well as the significantly less luminous afterglow of GRB\,120119A, which did show dust destruction at very early times.}
\label{KannPlot}
\end{center}
\end{figure}

The simplest SP model shows the GRB\,180325A 2175\,\AA\ bump is well fit by PAHs and silicates only. It is worth noting that silicates chemical bonds are weaker and are destroyed at low temperatures compared to graphite while the destruction times for both grains are not much different \citep{Evans94}. It is debated that the intense radiation of GRB could destroy dust, preferentially small grains, within a few pc of the burst \citep[e.g.,][]{Waxman00,Fruchter01,Perna03,Morgan14} and show a colour change within a few seconds of the burst related to dust destruction. This may explain why there is a lower contribution from PAHs in the dust model of GRB\,180325A, compared to dust models from within the Milky Way. \citet{Venemans01} suggested that a GRB peak optical flash with intensity less than $10^{48}$\,erg/s will destroy less dust along the line of sight.

Using data from \cite{Zafar18b}, \cite{Becerra21}, and as yet unpublished data sets (Kann et al., in prep., Jelinek et al., in prep.), we construct the optical afterglow light curve. Using the method of \cite{Kann06}, we correct the afterglow for the large host-galaxy extinction and shift it to a redshift $z=1$. We find an observed (corrected for Galactic extinction, $AB$ magnitudes) peak of the early flare of $r^\prime=16.30$\,mag \citep{Becerra21}, and a shift value of $dRc=-4.96$\,mag (see \citealt{Kann06} for more details), which implies $r^\prime=11.34$\,mag at $z=1$. Using WMAP cosmology \citep{Spergel03} we derive $m-M=43.96$ mag, and $M_{r^\prime}=-32.62$ mag, which is equivalent to a peak luminosity of the GRB afterglow of about 1.5$\times$10$^{51}$\,ergs/s, luminous enough that dust destruction may have taken place.

The afterglow light curve, in comparison to a large sample, is shown in Fig. \ref{KannPlot}. While the early flash is luminous, there are many GRBs with more luminous early afterglows. We point out that the afterglow of GRB\,120119A, the prototypical dust destruction detection \citep{Morgan14}, showed clear evidence of high extinction even in the later afterglow phase, so dust was destructed only partially. The early afterglow luminosity of this GRB is actually about one magnitude \emph{less} than that of GRB\,180325A, strengthening the case that the GRB\,180325A afterglow would have been capable of destroying dust in similar circumstances. However, we caution that dust destruction does not always take place, as in the case of GRB\,080607, a GRB with evidence for very high extinction \citep{Perley11} as well as molecules \citep{Prochaska09,Sheffer09}, which featured one of the most luminous early afterglows ever detected (\citealt{Perley11}, see Fig. \ref{KannPlot}), but showed no sign of any colour change in rapid $RJHK$ follow-up. This indicates that while bright prompt flashes may be capable of destroying dust, the actual occurrence is dependent on many factors, for example the distance of the dust screen to the GRB site within its host galaxy.

Dust grain models, including this work, use Mie Theory and assume perfectly spherical dust grains. However, dust grains are not perfect spheres, particularly as there can be shattering and sputtering of the dust grains \citep[e.g.,][]{Hirashita13}. Dust grains may have layers or a mantle, which would result in scattering different than Mie theory predicts \citep{Spitzer93,Draine03}. Future dust grain models should account for such asymmetries, particularly when modelling extinction curves from high energy events in environments beyond the Milky Way.

There is no clear detection of H$_2$ and CO molecules in the spectrum of GRB\,180325A \citep{Zafar18}. \citet{Draine02} modelled the photodissociation of H$_2$ and photoionisation of H$_2$, H, and He and destruction of dust grains by a GRB with strong optical-UV emission in a molecular cloud. They found that a GRB within a molecular cloud should show strong absorption lines at $1110$\,\AA\ $ < \lambda < 1705$\,\AA\ due to vibrational excitation of H$_2$. The energy of GRB 180325A indicates that it would destroy dust around it. We have not observed these strong H$_2$ absorption lines nor any other indicators of this GRB originating from within a molecular cloud. If that were observed, as this GRB has strong optical-UV emission, this dust destruction, photodissociation and photoionisation should be considered.

In this work we model a GRB curve with the dust grain-size distribution of WD2001 and \citet{Li01} and included contributions of iron for the first time using optical properties from \citet{Draine13}. 
We found that the contribution of iron nanoparticles was insignificant for the case of GRB\,180325A. There is potential for modelling more GRB curves and potentially extinction curves from other Galactic and extra-galactic sightlines. These models could contribute to the understanding of environments and comparisons between Galactic and distant environments. Further this can help in understanding the timescales of dust production and dust producers.

\section{Conclusions}
We modelled a GRB extinction curve with a dust grain distribution for the first time. We fit the observed extinction curve of the GRB\,180325A afterglow SED at $z=2.2486$ which exhibits a MW-type 2175\,\AA\ bump. We fit the observed extinction with the SPGI model for 110 sets of carbon abundance required in hydrocarbon molecules, $b_C$, and fraction of iron in free-flying nanoparticles, $Y_{\text{Fe}}$, in very small grains in addition to graphite and silicates components.
We fit for another three grain model permutations (SP, SPI, SPG) for 122 sets of $b_C$ and/or $Y_{\text{Fe}}$. We found that carbon and iron abundances for our case are degenerate and all four model permutations provide multiple best-fits for the GRB\,180325A extinction curve. We found that the contribution of iron nanoparticles to the extinction of GRB\,180325A is insignificant. Furthermore, we find that the inclusion of graphite does not significantly improve the fit of the 2175 {\AA} bump. Due to insignificant improvement in the fit when using the complex models, we favour the simplest model of silicates and PAHs to describe the extinction curve of GRB\,180325A.

The carbon abundance is found to be degenerate and the afterglow SED of GRB\,180325A can be described by the carbon abundance range $(0.0 \leq b_C \leq 0.7) \times 10^{-5}$.

The silicates' contribution dominates the entire wavelength range of the GRB extinction curve and graphite contributes to the bump and UV region. We conclude that the observed extinction is produced by a mixture of different grain-sizes and compositions where each component provided a significant contribution. However, none of the models fit the steep UV extinction well. Potentially there are other types of dust grains that are contributing to that part of the extinction. Further research into other dust grain compositions and size distributions may lead to better extinction curve fits and a better understanding of dusty environments.

\section*{Acknowledgements}
We would like to thank Brandon Hensley for sharing optical properties of metallic iron and Bruce Draine for useful and constructive comments. DAK acknowledges support from Spanish National Research Project RTI2018-098104-J-I00 (GRBPhot), and thanks M. Blazek for calculations. Thanks to the anonymous referee for constructive comments to improve the work.

\section*{Data Availability}
The data underlying this article are partly available at https://www.astro.princeton.edu/$\sim$draine/. The iron nanoparticle data underlying this article cannot be shared publicly as the original authors do not give open access.



\bibliographystyle{mnras}
\bibliography{dust-model} 




\appendix

\section{Additional Results}
\begin{figure}
\begin{center}
{\includegraphics[width=\columnwidth,bb=35 85 915 620]{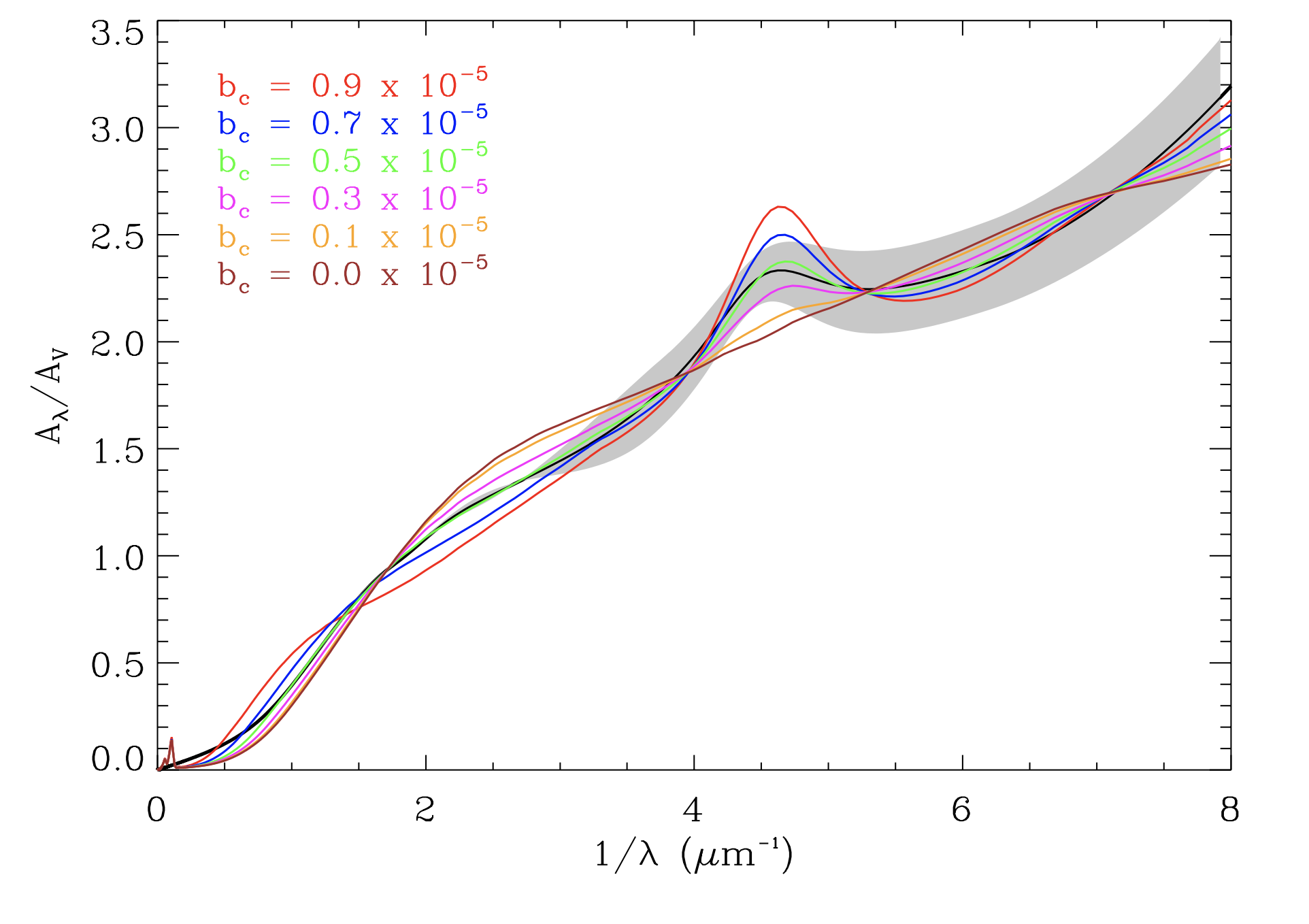}}
\caption{Modelled extinction curves overlaid on observed extinction curve (solid black line). The gray-shaded area represents the 1$\sigma$ uncertainty on the observed extinction curve. The modelled extinction curves for the SP model with $b_C=(0.0$, 0.1, 0.3, 0.5, 0.7, 0.9)\,$\times10^{-5}$ are shown to present low and high $b_C$ values around the best-fit.}
\label{fig:multidustsp}
\end{center}
\end{figure}

\begin{figure}
\begin{center}
{\includegraphics[width=\columnwidth,bb=35 85 915 620]{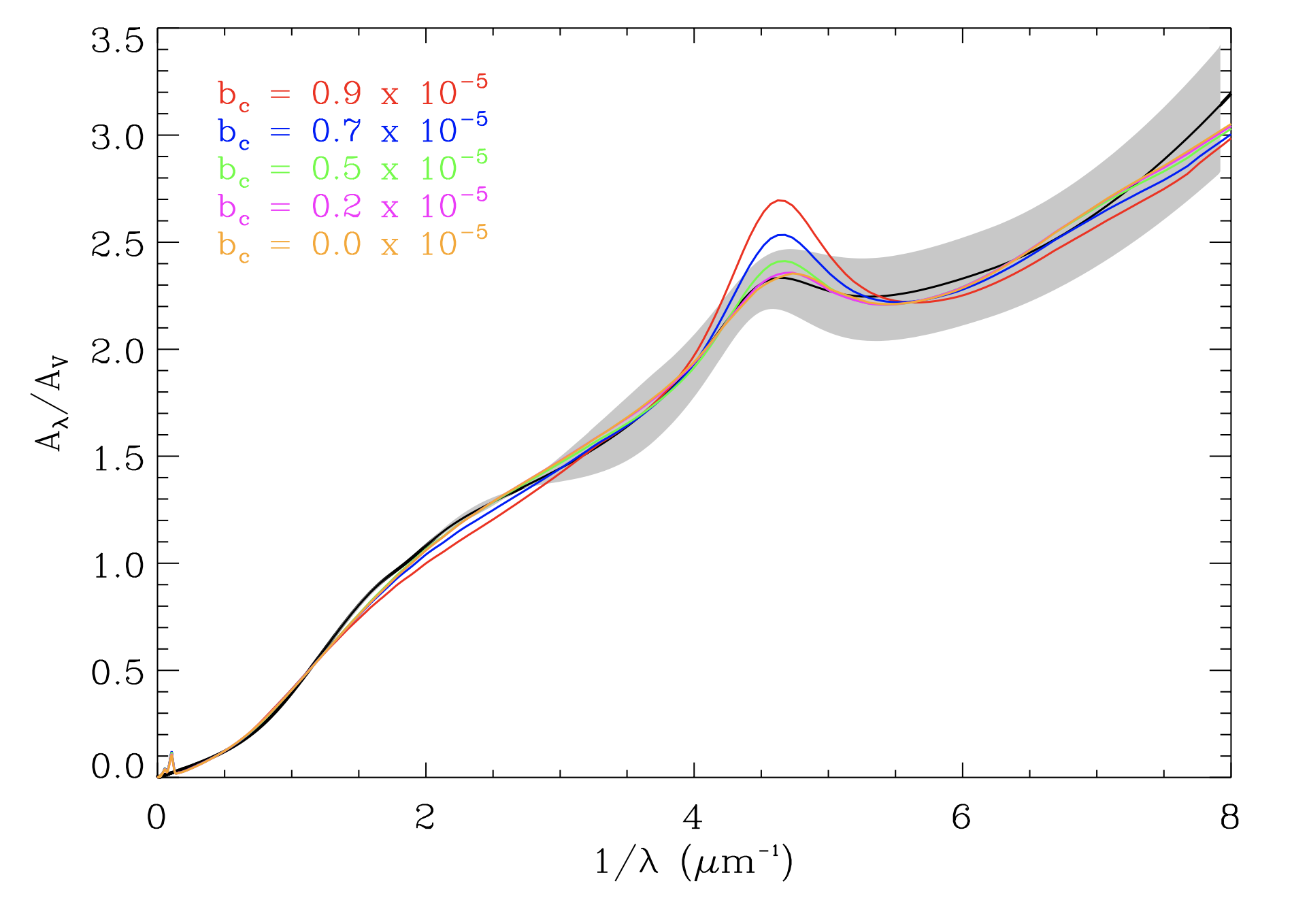}}
\caption{Modelled extinction curves overlaid on observed extinction curve (solid black line). The gray-shaded area represents the 1$\sigma$ uncertainty on the observed extinction curve. The modelled extinction curves for the SPG model with $b_C=(0.0$, 0.2, 0.5, 0.7, 0.9)\,$\times10^{-5}$ are shown to present low and high $b_C$ values around the best-fit.}
\label{fig:multidustspg}
\end{center}
\end{figure}

\begin{figure}
\begin{center}
{\includegraphics[width=\columnwidth,bb=30 85 915 650]{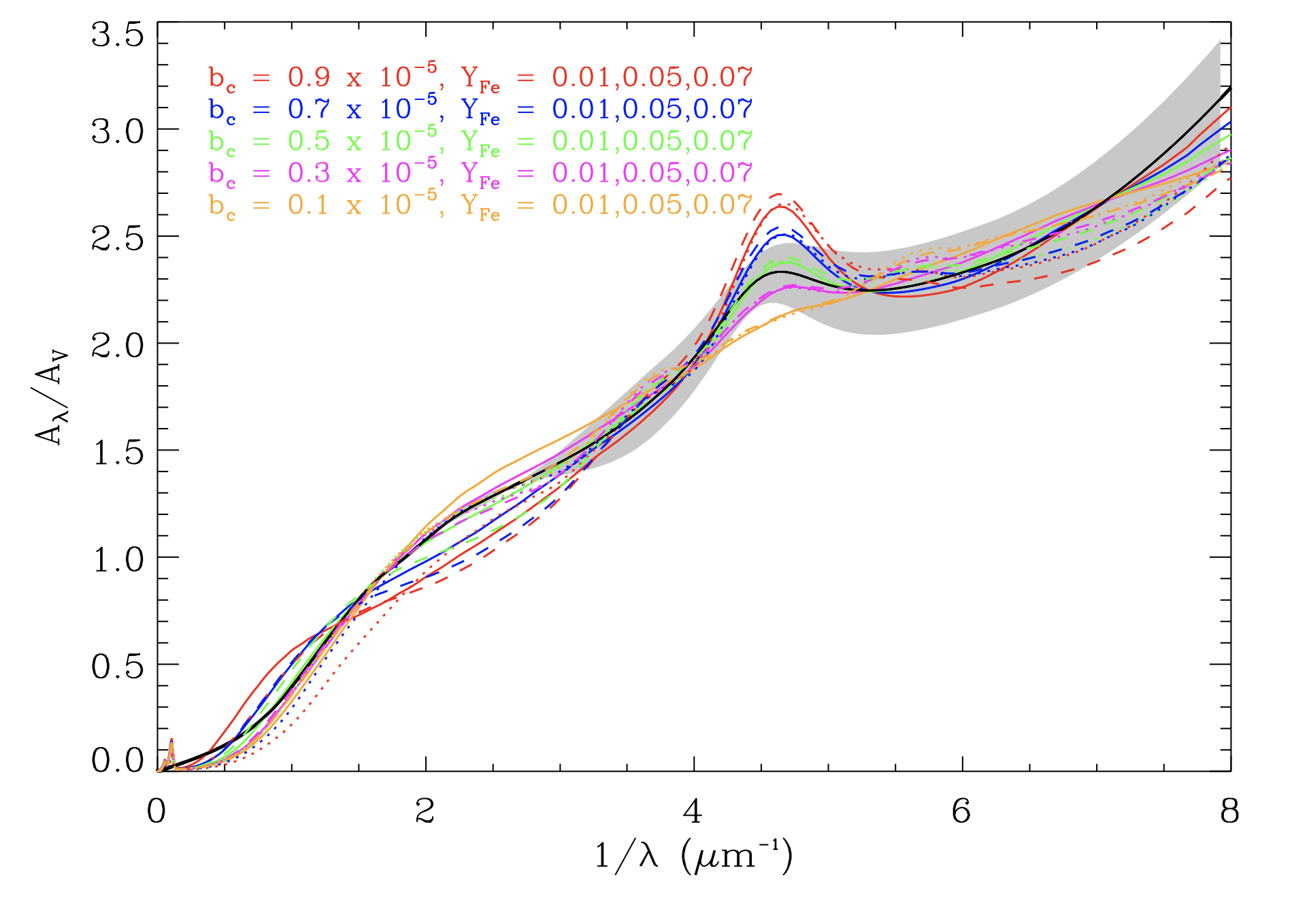}}
\caption{Modelled extinction curves overlaid on observed extinction curve (solid black line). The gray-shaded area represents the 1$\sigma$ uncertainty on the observed extinction curve. The modelled extinction curves for the SPI model with $b_C=(0.1$, 0.3, 0.5, 0.7, 0.9)\,$\times10^{-5}$ and $Y_{\text{Fe}}$ = 0.01, 0.05, 0.07 (solid, dashed, dotted, respectively) are shown to present low and high $b_C$ and $Y_{\text{Fe}}$ values around the best-fit.}
\label{fig:multidustspi}
\end{center}
\end{figure}

\begin{figure}
\begin{center}
{\includegraphics[width=\columnwidth,bb=30 85 915 620]{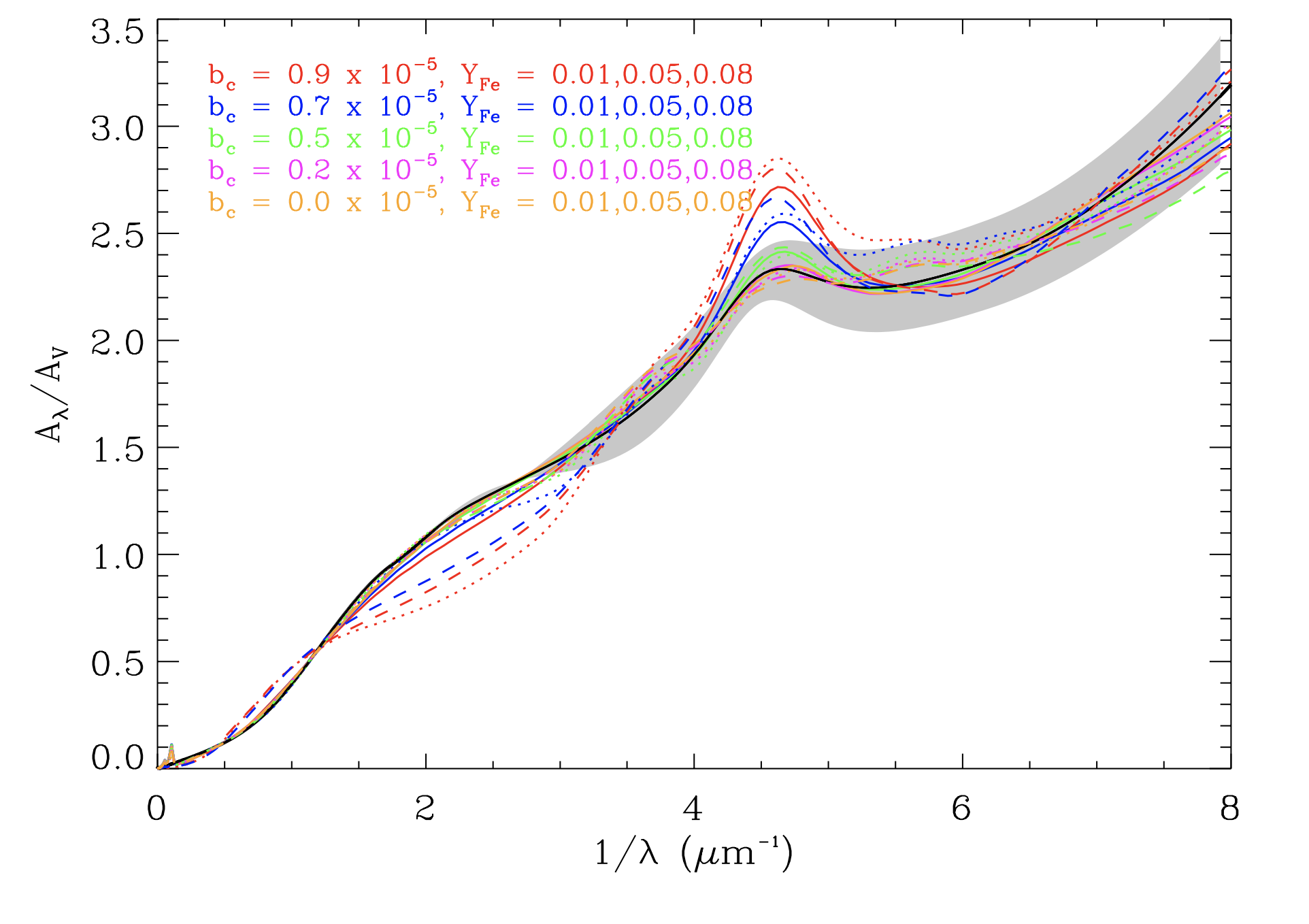}}
\caption{Modelled extinction curves overlaid on observed extinction curve (solid black line). The gray-shaded area represents the 1$\sigma$ uncertainty on the observed extinction curve. The modelled extinction curves for the SPGI model with $b_C=(0.0$, 0.2, 0.5, 0.7, 0.9)\,$\times10^{-5}$ and $Y_{\text{Fe}}$ = 0.01, 0.05, 0.08 (solid, dashed, dotted, respectively) are shown to present low and high $b_C$ and $Y_{\text{Fe}}$ values around the best-fit.}
\label{fig:multidustspgi}
\end{center}
\end{figure}

In order to illustrate the degeneracy in the fit of $b_C$ in all four model permutations and $Y_\text{Fe}$ in two models, we have included figures \ref{fig:multidustsp}, \ref{fig:multidustspg}, \ref{fig:multidustspi} and \ref{fig:multidustspgi} for model SP, SPG, SPI and SPGI, respectively. In these figures, we have included a selection of $b_C$ and $Y_\text{Fe}$ values to show how the behaviour of the models changes across the range of possible values.

We have also included Table \ref{table:2} which provides the optimal parameters for each model permutation along with the $\chi_\nu$ and p-value. This demonstrates how the fits generally get worse at values of $b_C > 0.7 \times 10^{-5}$.

\begin{table*}
\caption{Fitted grain-size distribution parameter values for GRB\,180325A. For each model, $\chi^2_\nu$ and p-values are provided.
The number of degrees of freedom for each model is included along with the adopted name.}
\label{table:2} 
\centering     
\setlength{\tabcolsep}{6pt}
\begin{tabular}{c c c c c c c c c c c c c }  
\hline\hline                        
$b_C$ & $Y_{\text{Fe}}$ & $\alpha_g$ & $\beta_g$ & $a_{t,g}$ & $a_{c,g}$ & $C_g$ & $\alpha_s$ & $\beta_s$ & $a_{t,s}$ & $C_s$ & $\chi^2_\nu$ & p-value\\
$10^{-5}$ &  &  &  & $\mu$m & $\mu$m & $10^{-16}$ &  &  & $\mu$m & $10^{-16}$ &  &  \\
\hline
\textbf{SP} (781)& \\
0.0 & $\cdots$ & $\cdots$ & $\cdots$ & $\cdots$ & $\cdots$ & $\cdots$ & -2.85 & 31.09 & 0.19 & 7.41 & 6.13 & 0.00 \\
0.1 & $\cdots$ & $\cdots$ & $\cdots$ & $\cdots$ & $\cdots$ & $\cdots$ & -2.86 & 28.25 & 0.20 & 6.73 & 4.30 & 0.00 \\
0.2 & $\cdots$ & $\cdots$ & $\cdots$ & $\cdots$ & $\cdots$ & $\cdots$ & -2.88 & 25.82 & 0.21 & 5.98 & 2.89 & 0.00 \\
0.3 & $\cdots$ & $\cdots$ & $\cdots$ & $\cdots$ & $\cdots$ & $\cdots$ & -2.89 & 22.92 & 0.23 & 5.35 & 1.99 & 0.00 \\
0.4 & $\cdots$ & $\cdots$ & $\cdots$ & $\cdots$ & $\cdots$ & $\cdots$ & -2.89 & 19.49 & 0.25 & 4.83 & 1.48 & 0.00 \\
0.5 & $\cdots$ & $\cdots$ & $\cdots$ & $\cdots$ & $\cdots$ & $\cdots$ & -2.83 & 12.11 & 0.27 & 6.04 & 1.06 & 0.11 \\
0.6 & $\cdots$ & $\cdots$ & $\cdots$ & $\cdots$ & $\cdots$ & $\cdots$ & -2.74 & 6.85 & 0.30 & 7.79 & 1.52 & 0.00 \\
0.7 & $\cdots$ & $\cdots$ & $\cdots$ & $\cdots$ & $\cdots$ & $\cdots$ & -2.77 & 7.24 & 0.34 & 4.93 & 1.90 & 0.00 \\
0.8 & $\cdots$ & $\cdots$ & $\cdots$ & $\cdots$ & $\cdots$ & $\cdots$ & -2.81 & 8.00 & 0.40 & 2.94 & 2.82 & 0.00 \\
0.9 & $\cdots$ & $\cdots$ & $\cdots$ & $\cdots$ & $\cdots$ & $\cdots$ & -2.85 & 9.28 & 0.48 & 1.53 & 4.31 & 0.00 \\
1.0 & $\cdots$ & $\cdots$ & $\cdots$ & $\cdots$ & $\cdots$ & $\cdots$ & -2.90 & 11.96 & 0.60 & 0.66 & 6.24 & 0.00 \\
\hline
\textbf{SPI} (781)& \\
0.1 & 0.01 & $\cdots$ & $\cdots$ & $\cdots$ & $\cdots$ & $\cdots$ & -2.80 & 26.11 & 0.21 & 6.59 & 4.03 & 0.00 \\
0.1 & 0.02 & $\cdots$ & $\cdots$ & $\cdots$ & $\cdots$ & $\cdots$ & -2.73 & 24.30 & 0.21 & 6.55 & 3.86 & 0.00 \\
0.1 & 0.03 & $\cdots$ & $\cdots$ & $\cdots$ & $\cdots$ & $\cdots$ & -2.65 & 22.92 & 0.21 & 6.58 & 3.78 & 0.00 \\
0.1 & 0.04 & $\cdots$ & $\cdots$ & $\cdots$ & $\cdots$ & $\cdots$ & -2.53 & 21.59 & 0.22 & 6.86 & 3.80 & 0.00 \\
0.1 & 0.05 & $\cdots$ & $\cdots$ & $\cdots$ & $\cdots$ & $\cdots$ & -2.38 & 20.11 & 0.21 & 7.55 & 3.88 & 0.00 \\
0.1 & 0.06 & $\cdots$ & $\cdots$ & $\cdots$ & $\cdots$ & $\cdots$ & -2.15 & 19.26 & 0.21 & 8.48 & 4.01 & 0.00 \\
0.1 & 0.07 & $\cdots$ & $\cdots$ & $\cdots$ & $\cdots$ & $\cdots$ & -1.80 & 18.58 & 0.20 & 9.83 & 4.20 & 0.00 \\
0.1 & 0.08 & $\cdots$ & $\cdots$ & $\cdots$ & $\cdots$ & $\cdots$ & -1.23 & 16.10 & 0.18 & 12.61 & 4.61 & 0.00 \\
0.1 & 0.09 & $\cdots$ & $\cdots$ & $\cdots$ & $\cdots$ & $\cdots$ & -0.39 & 18.50 & 0.17 & 11.06 & 5.74 & 0.00 \\
0.1 & 0.10 & $\cdots$ & $\cdots$ & $\cdots$ & $\cdots$ & $\cdots$ & 0.74 & 24.67 & 0.15 & 6.54 & 8.41 & 0.00 \\
0.2 & 0.01 & $\cdots$ & $\cdots$ & $\cdots$ & $\cdots$ & $\cdots$ & -2.82 & 24.02 & 0.22 & 5.79 & 2.68 & 0.00 \\
0.2 & 0.02 & $\cdots$ & $\cdots$ & $\cdots$ & $\cdots$ & $\cdots$ & -2.75 & 22.13 & 0.23 & 5.79 & 2.58 & 0.00 \\
0.2 & 0.03 & $\cdots$ & $\cdots$ & $\cdots$ & $\cdots$ & $\cdots$ & -2.65 & 18.72 & 0.23 & 6.54 & 2.58 & 0.00 \\
0.2 & 0.04 & $\cdots$ & $\cdots$ & $\cdots$ & $\cdots$ & $\cdots$ & -2.45 & 10.78 & 0.23 & 11.41 & 2.65 & 0.00 \\
0.2 & 0.05 & $\cdots$ & $\cdots$ & $\cdots$ & $\cdots$ & $\cdots$ & -2.23 & 7.64 & 0.22 & 16.61 & 2.77 & 0.00 \\
0.2 & 0.06 & $\cdots$ & $\cdots$ & $\cdots$ & $\cdots$ & $\cdots$ & -1.92 & 5.60 & 0.21 & 24.45 & 2.91 & 0.00 \\
0.2 & 0.07 & $\cdots$ & $\cdots$ & $\cdots$ & $\cdots$ & $\cdots$ & -1.66 & 15.08 & 0.20 & 11.47 & 3.04 & 0.00 \\
0.2 & 0.08 & $\cdots$ & $\cdots$ & $\cdots$ & $\cdots$ & $\cdots$ & -0.94 & 21.55 & 0.18 & 9.42 & 3.31 & 0.00 \\
0.2 & 0.09 & $\cdots$ & $\cdots$ & $\cdots$ & $\cdots$ & $\cdots$ & 0.24 & 19.09 & 0.16 & 9.67 & 4.48 & 0.00 \\
0.2 & 0.10 & $\cdots$ & $\cdots$ & $\cdots$ & $\cdots$ & $\cdots$ & 1.89 & 16.60 & 0.13 & 6.18 & 7.52 & 0.00 \\
0.3 & 0.01 & $\cdots$ & $\cdots$ & $\cdots$ & $\cdots$ & $\cdots$ & -2.83 & 20.95 & 0.24 & 5.19 & 1.75 & 0.00 \\
0.3 & 0.02 & $\cdots$ & $\cdots$ & $\cdots$ & $\cdots$ & $\cdots$ & -2.70 & 13.74 & 0.24 & 7.44 & 1.72 & 0.00 \\
0.3 & 0.03 & $\cdots$ & $\cdots$ & $\cdots$ & $\cdots$ & $\cdots$ & -2.53 & 8.07 & 0.24 & 12.07 & 1.78 & 0.00 \\
0.3 & 0.04 & $\cdots$ & $\cdots$ & $\cdots$ & $\cdots$ & $\cdots$ & -2.36 & 6.07 & 0.24 & 15.66 & 1.92 & 0.00 \\
0.3 & 0.05 & $\cdots$ & $\cdots$ & $\cdots$ & $\cdots$ & $\cdots$ & -2.14 & 4.84 & 0.23 & 20.69 & 2.11 & 0.00 \\
0.3 & 0.06 & $\cdots$ & $\cdots$ & $\cdots$ & $\cdots$ & $\cdots$ & -1.80 & 3.70 & 0.22 & 30.98 & 2.26 & 0.00 \\
0.3 & 0.07 & $\cdots$ & $\cdots$ & $\cdots$ & $\cdots$ & $\cdots$ & -1.23 & 3.31 & 0.20 & 42.77 & 2.30 & 0.00 \\
0.3 & 0.08 & $\cdots$ & $\cdots$ & $\cdots$ & $\cdots$ & $\cdots$ & -0.41 & 16.12 & 0.17 & 12.18 & 2.43 & 0.00 \\
0.3 & 0.09 & $\cdots$ & $\cdots$ & $\cdots$ & $\cdots$ & $\cdots$ & 1.26 & 16.53 & 0.14 & 8.31 & 3.72 & 0.00 \\
0.3 & 0.10 & $\cdots$ & $\cdots$ & $\cdots$ & $\cdots$ & $\cdots$ & 3.76 & 13.09 & 0.11 & 2.52 & 7.29 & 0.00 \\
0.4 & 0.01 & $\cdots$ & $\cdots$ & $\cdots$ & $\cdots$ & $\cdots$ & -2.77 & 12.69 & 0.26 & 6.70 & 1.28 & 0.00 \\
0.4 & 0.02 & $\cdots$ & $\cdots$ & $\cdots$ & $\cdots$ & $\cdots$ & -2.63 & 7.98 & 0.27 & 9.58 & 1.26 & 0.00 \\
0.4 & 0.03 & $\cdots$ & $\cdots$ & $\cdots$ & $\cdots$ & $\cdots$ & -2.52 & 6.92 & 0.27 & 9.96 & 1.41 & 0.00 \\
0.4 & 0.04 & $\cdots$ & $\cdots$ & $\cdots$ & $\cdots$ & $\cdots$ & -2.40 & 6.27 & 0.28 & 10.65 & 1.64 & 0.00 \\
0.4 & 0.05 & $\cdots$ & $\cdots$ & $\cdots$ & $\cdots$ & $\cdots$ & -2.17 & 5.07 & 0.26 & 14.93 & 1.93 & 0.00 \\
0.4 & 0.06 & $\cdots$ & $\cdots$ & $\cdots$ & $\cdots$ & $\cdots$ & -1.75 & 3.61 & 0.23 & 27.93 & 2.10 & 0.00 \\
0.4 & 0.07 & $\cdots$ & $\cdots$ & $\cdots$ & $\cdots$ & $\cdots$ & -0.87 & 2.44 & 0.19 & 54.50 & 1.97 & 0.00 \\
0.4 & 0.08 & $\cdots$ & $\cdots$ & $\cdots$ & $\cdots$ & $\cdots$ & 0.61 & 6.70 & 0.15 & 23.76 & 1.94 & 0.00 \\
0.4 & 0.09 & $\cdots$ & $\cdots$ & $\cdots$ & $\cdots$ & $\cdots$ & 3.68 & 0.49 & 0.12 & 38.27 & 3.55 & 0.00 \\
0.4 & 0.10 & $\cdots$ & $\cdots$ & $\cdots$ & $\cdots$ & $\cdots$ & 8.64 & -5.46 & 0.09 & 11.40 & 7.79 & 0.00 \\
0.5 & 0.01 & $\cdots$ & $\cdots$ & $\cdots$ & $\cdots$ & $\cdots$ & -2.70 & 7.72 & 0.28 & 8.08 & 1.05 & 0.17 \\
0.5 & 0.02 & $\cdots$ & $\cdots$ & $\cdots$ & $\cdots$ & $\cdots$ & -2.64 & 7.54 & 0.31 & 6.84 & 1.20 & 0.00 \\
0.5 & 0.03 & $\cdots$ & $\cdots$ & $\cdots$ & $\cdots$ & $\cdots$ & -2.57 & 7.55 & 0.32 & 5.89 & 1.45 & 0.00 \\
0.5 & 0.04 & $\cdots$ & $\cdots$ & $\cdots$ & $\cdots$ & $\cdots$ & -2.45 & 7.11 & 0.33 & 6.05 & 1.80 & 0.00 \\
\hline 
\end{tabular}
\end{table*}
\begin{table*}
\contcaption{}
\centering     
\setlength{\tabcolsep}{6pt}
\begin{tabular}{c c c c c c c c c c c c c }  
\hline\hline                        
$b_C$ & $Y_{\text{Fe}}$ & $\alpha_g$ & $\beta_g$ & $a_{t,g}$ & $a_{c,g}$ & $C_g$ & $\alpha_s$ & $\beta_s$ & $a_{t,s}$ & $C_s$ & $\chi^2_\nu$ & p-value\\
$10^{-5}$ &  &  &  & $\mu$m & $\mu$m & $10^{-16}$ &  &  & $\mu$m & $10^{-16}$ &  &  \\
\hline
\textbf{SPI} (776)& \\
0.5 & 0.05 & $\cdots$ & $\cdots$ & $\cdots$ & $\cdots$ & $\cdots$ & -2.24 & 6.04 & 0.31 & 8.36 & 2.20 & 0.00 \\
0.5 & 0.06 & $\cdots$ & $\cdots$ & $\cdots$ & $\cdots$ & $\cdots$ & -1.69 & 3.72 & 0.24 & 24.07 & 2.42 & 0.00 \\
0.5 & 0.07 & $\cdots$ & $\cdots$ & $\cdots$ & $\cdots$ & $\cdots$ & -0.23 & 2.38 & 0.18 & 58.43 & 2.00 & 0.00 \\
0.5 & 0.08 & $\cdots$ & $\cdots$ & $\cdots$ & $\cdots$ & $\cdots$ & 2.79 & 0.85 & 0.13 & 50.35 & 1.88 & 0.00 \\
0.5 & 0.09 & $\cdots$ & $\cdots$ & $\cdots$ & $\cdots$ & $\cdots$ & 7.86 & -9.03 & 0.09 & 46.47 & 4.05 & 0.00 \\
0.5 & 0.10 & $\cdots$ & $\cdots$ & $\cdots$ & $\cdots$ & $\cdots$ & 11.58 & -26.13 & 0.06 & 0.43 & 9.08 & 0.00 \\
0.6 & 0.01 & $\cdots$ & $\cdots$ & $\cdots$ & $\cdots$ & $\cdots$ & -2.72 & 7.58 & 0.33 & 5.52 & 1.39 & 0.00 \\
0.6 & 0.02 & $\cdots$ & $\cdots$ & $\cdots$ & $\cdots$ & $\cdots$ & -2.67 & 7.87 & 0.36 & 4.33 & 1.55 & 0.00 \\
0.6 & 0.03 & $\cdots$ & $\cdots$ & $\cdots$ & $\cdots$ & $\cdots$ & -2.58 & 7.34 & 0.37 & 4.22 & 1.95 & 0.00 \\
0.6 & 0.04 & $\cdots$ & $\cdots$ & $\cdots$ & $\cdots$ & $\cdots$ & -2.46 & 7.05 & 0.37 & 4.47 & 2.45 & 0.00 \\
0.6 & 0.05 & $\cdots$ & $\cdots$ & $\cdots$ & $\cdots$ & $\cdots$ & -2.25 & 6.39 & 0.35 & 5.93 & 2.96 & 0.00 \\
0.6 & 0.06 & $\cdots$ & $\cdots$ & $\cdots$ & $\cdots$ & $\cdots$ & -1.56 & 3.61 & 0.25 & 23.25 & 3.21 & 0.00 \\
0.6 & 0.07 & $\cdots$ & $\cdots$ & $\cdots$ & $\cdots$ & $\cdots$ & 2.02 & 0.17 & 0.14 & 132.25 & 2.26 & 0.00 \\
0.6 & 0.08 & $\cdots$ & $\cdots$ & $\cdots$ & $\cdots$ & $\cdots$ & 7.17 & -8.72 & 0.10 & 96.00 & 2.40 & 0.00 \\
0.6 & 0.09 & $\cdots$ & $\cdots$ & $\cdots$ & $\cdots$ & $\cdots$ & 11.64 & -30.14 & 0.06 & 0.41 & 5.31 & 0.00 \\
0.6 & 0.10 & $\cdots$ & $\cdots$ & $\cdots$ & $\cdots$ & $\cdots$ & 11.96 & -72.98 & 0.06 & 0.35 & 11.26 & 0.00 \\
0.7 & 0.01 & $\cdots$ & $\cdots$ & $\cdots$ & $\cdots$ & $\cdots$ & -2.75 & 8.03 & 0.38 & 3.46 & 2.00 & 0.00 \\
0.7 & 0.02 & $\cdots$ & $\cdots$ & $\cdots$ & $\cdots$ & $\cdots$ & -2.70 & 8.28 & 0.41 & 2.86 & 2.40 & 0.00 \\
0.7 & 0.03 & $\cdots$ & $\cdots$ & $\cdots$ & $\cdots$ & $\cdots$ & -2.63 & 8.45 & 0.42 & 2.61 & 2.97 & 0.00 \\
0.7 & 0.04 & $\cdots$ & $\cdots$ & $\cdots$ & $\cdots$ & $\cdots$ & -2.48 & 7.28 & 0.41 & 3.31 & 3.64 & 0.00 \\
0.7 & 0.05 & $\cdots$ & $\cdots$ & $\cdots$ & $\cdots$ & $\cdots$ & -2.24 & 6.41 & 0.37 & 4.90 & 4.27 & 0.00 \\
0.7 & 0.06 & $\cdots$ & $\cdots$ & $\cdots$ & $\cdots$ & $\cdots$ & -1.01 & 2.88 & 0.22 & 36.36 & 4.40 & 0.00 \\
0.7 & 0.07 & $\cdots$ & $\cdots$ & $\cdots$ & $\cdots$ & $\cdots$ & 5.94 & -0.89 & 0.10 & 43.59 & 2.85 & 0.00 \\
0.7 & 0.08 & $\cdots$ & $\cdots$ & $\cdots$ & $\cdots$ & $\cdots$ & 11.62 & -45.40 & 0.06 & 0.60 & 3.64 & 0.00 \\
0.7 & 0.09 & $\cdots$ & $\cdots$ & $\cdots$ & $\cdots$ & $\cdots$ & 12.08 & -87.95 & 0.05 & 0.29 & 7.44 & 0.00 \\
0.7 & 0.10 & $\cdots$ & $\cdots$ & $\cdots$ & $\cdots$ & $\cdots$ & 13.97 & -188.86 & 0.04 & 0.00 & 14.29 & 0.00 \\
0.8 & 0.01 & $\cdots$ & $\cdots$ & $\cdots$ & $\cdots$ & $\cdots$ & -2.79 & 9.29 & 0.45 & 1.92 & 3.18 & 0.00 \\
0.8 & 0.02 & $\cdots$ & $\cdots$ & $\cdots$ & $\cdots$ & $\cdots$ & -2.75 & 10.10 & 0.51 & 1.31 & 3.74 & 0.00 \\
0.8 & 0.03 & $\cdots$ & $\cdots$ & $\cdots$ & $\cdots$ & $\cdots$ & -2.71 & 11.40 & 0.55 & 0.94 & 4.48 & 0.00 \\
0.8 & 0.04 & $\cdots$ & $\cdots$ & $\cdots$ & $\cdots$ & $\cdots$ & -2.57 & 9.92 & 0.54 & 1.19 & 5.35 & 0.00 \\
0.8 & 0.05 & $\cdots$ & $\cdots$ & $\cdots$ & $\cdots$ & $\cdots$ & -2.20 & 6.07 & 0.39 & 4.51 & 6.13 & 0.00 \\
0.8 & 0.06 & $\cdots$ & $\cdots$ & $\cdots$ & $\cdots$ & $\cdots$ & 5.17 & -2.51 & 0.11 & 182.25 & 5.37 & 0.00 \\
0.8 & 0.07 & $\cdots$ & $\cdots$ & $\cdots$ & $\cdots$ & $\cdots$ & 11.33 & -59.52 & 0.06 & 1.29 & 4.06 & 0.00 \\
0.8 & 0.08 & $\cdots$ & $\cdots$ & $\cdots$ & $\cdots$ & $\cdots$ & 12.38 & -121.10 & 0.05 & 0.19 & 5.71 & 0.00 \\
0.8 & 0.09 & $\cdots$ & $\cdots$ & $\cdots$ & $\cdots$ & $\cdots$ & 14.49 & -248.13 & 0.04 & 0.00 & 10.41 & 0.00 \\
0.8 & 0.10 & $\cdots$ & $\cdots$ & $\cdots$ & $\cdots$ & $\cdots$ & 13.82 & -290.43 & 0.04 & 0.00 & 18.27 & 0.00 \\
0.9 & 0.01 & $\cdots$ & $\cdots$ & $\cdots$ & $\cdots$ & $\cdots$ & -2.85 & 11.98 & 0.57 & 0.79 & 4.81 & 0.00 \\
0.9 & 0.02 & $\cdots$ & $\cdots$ & $\cdots$ & $\cdots$ & $\cdots$ & -2.81 & 12.62 & 0.62 & 0.61 & 5.54 & 0.00 \\
0.9 & 0.03 & $\cdots$ & $\cdots$ & $\cdots$ & $\cdots$ & $\cdots$ & -2.76 & 13.69 & 0.68 & 0.45 & 6.47 & 0.00 \\
0.9 & 0.04 & $\cdots$ & $\cdots$ & $\cdots$ & $\cdots$ & $\cdots$ & -2.60 & 10.60 & 0.62 & 0.77 & 7.56 & 0.00 \\
0.9 & 0.05 & $\cdots$ & $\cdots$ & $\cdots$ & $\cdots$ & $\cdots$ & -0.43 & -856.66 & 0.39 & 26032.70 & 9.21 & 0.00 \\
0.9 & 0.06 & $\cdots$ & $\cdots$ & $\cdots$ & $\cdots$ & $\cdots$ & 10.16 & -94.27 & 0.07 & 17.17 & 6.58 & 0.00 \\
0.9 & 0.07 & $\cdots$ & $\cdots$ & $\cdots$ & $\cdots$ & $\cdots$ & 14.20 & -147.66 & 0.04 & 0.00 & 6.05 & 0.00 \\
0.9 & 0.08 & $\cdots$ & $\cdots$ & $\cdots$ & $\cdots$ & $\cdots$ & 14.52 & -249.33 & 0.04 & 0.00 & 8.63 & 0.00 \\
0.9 & 0.09 & $\cdots$ & $\cdots$ & $\cdots$ & $\cdots$ & $\cdots$ & 0.00 & -149.81 & 1.44 & 240.49 & 19.64 & 0.00 \\
0.9 & 0.10 & $\cdots$ & $\cdots$ & $\cdots$ & $\cdots$ & $\cdots$ & 13.84 & -335.76 & 0.04 & 0.00 & 23.16 & 0.00 \\
1.0 & 0.01 & $\cdots$ & $\cdots$ & $\cdots$ & $\cdots$ & $\cdots$ & -2.90 & 14.81 & 0.72 & 0.34 & 6.91 & 0.00 \\
1.0 & 0.02 & $\cdots$ & $\cdots$ & $\cdots$ & $\cdots$ & $\cdots$ & -2.87 & 16.72 & 0.81 & 0.22 & 7.82 & 0.00 \\
1.0 & 0.03 & $\cdots$ & $\cdots$ & $\cdots$ & $\cdots$ & $\cdots$ & -2.81 & 17.20 & 0.86 & 0.19 & 8.96 & 0.00 \\
1.0 & 0.04 & $\cdots$ & $\cdots$ & $\cdots$ & $\cdots$ & $\cdots$ & -1.08 & -12724.80 & 1.03 & 22943.80 & 11.58 & 0.00 \\
1.0 & 0.05 & $\cdots$ & $\cdots$ & $\cdots$ & $\cdots$ & $\cdots$ & 9.80 & -28243.20 & 0.07 & 9262.12 & 11.19 & 0.00 \\
1.0 & 0.06 & $\cdots$ & $\cdots$ & $\cdots$ & $\cdots$ & $\cdots$ & 12.25 & -151.90 & 0.05 & 0.27 & 8.55 & 0.00 \\
1.0 & 0.07 & $\cdots$ & $\cdots$ & $\cdots$ & $\cdots$ & $\cdots$ & 14.48 & -252.10 & 0.03 & 0.00 & 8.95 & 0.00 \\
1.0 & 0.08 & $\cdots$ & $\cdots$ & $\cdots$ & $\cdots$ & $\cdots$ & 0.02 & -127.60 & 1.39 & 251.38 & 18.14 & 0.00 \\
1.0 & 0.09 & $\cdots$ & $\cdots$ & $\cdots$ & $\cdots$ & $\cdots$ & 13.73 & -356.29 & 0.04 & 0.00 & 19.20 & 0.00 \\
1.0 & 0.10 & $\cdots$ & $\cdots$ & $\cdots$ & $\cdots$ & $\cdots$ & 9.14 & -392.11 & 0.07 & 91.41 & 29.17 & 0.00 \\
\hline
\textbf{SPG} (776)& \\
0.0 & $\cdots$ & -2.21 & -0.87 & 0.10 & 0.78 & 256.52 & -3.27 & 130.29 & 0.24 & 0.52 & 1.18 & 0.00 \\
0.1 & $\cdots$ & -2.12 & -0.92 & 0.10 & 0.75 & 229.83 & -3.23 & 97.86 & 0.24 & 0.72 & 1.06 & 0.11 \\
0.2 & $\cdots$ & -2.00 & -0.92 & 0.09 & 0.71 & 268.98 & -3.20 & 72.10 & 0.24 & 1.01 & 1.03 & 0.29 \\
0.3 & $\cdots$ & -1.85 & -0.94 & 0.09 & 0.68 & 291.14 & -3.13 & 41.80 & 0.24 & 1.84 & 1.07 & 0.10 \\
\hline 
\end{tabular}
\end{table*}
\begin{table*}
\contcaption{}
\centering     
\setlength{\tabcolsep}{6pt}
\begin{tabular}{c c c c c c c c c c c c c }  
\hline\hline                        
$b_C$ & $Y_{\text{Fe}}$ & $\alpha_g$ & $\beta_g$ & $a_{t,g}$ & $a_{c,g}$ & $C_g$ & $\alpha_s$ & $\beta_s$ & $a_{t,s}$ & $C_s$ & $\chi^2_\nu$ & p-value\\
$10^{-5}$ &  &  &  & $\mu$m & $\mu$m & $10^{-16}$ &  &  & $\mu$m & $10^{-16}$ &  &  \\
\hline
\textbf{SPG} (776)& \\
0.4 & $\cdots$ & -1.68 & -1.03 & 0.08 & 0.66 & 271.71 & -2.53 & 1.51 & 0.23 & 36.86 & 1.10 & 0.03 \\
0.5 & $\cdots$ & -1.39 & -0.82 & 0.07 & 0.62 & 260.09 & -2.61 & 2.51 & 0.23 & 27.33 & 1.17 & 0.00 \\
0.6 & $\cdots$ & -1.00 & -0.58 & 0.05 & 0.60 & 171.87 & -2.67 & 3.84 & 0.24 & 18.82 & 1.38 & 0.00 \\
0.7 & $\cdots$ & -0.66 & -0.35 & 0.03 & 0.60 & 127.52 & -2.69 & 4.45 & 0.25 & 14.75 & 1.66 & 0.00 \\
0.8 & $\cdots$ & -0.49 & -0.91 & 0.09 & 0.58 & 39.72 & -2.70 & 4.86 & 0.26 & 11.91 & 2.89 & 0.00 \\
0.9 & $\cdots$ & -0.37 & -0.96 & 0.10 & 0.57 & 27.48 & -2.71 & 5.11 & 0.28 & 9.94 & 3.97 & 0.00 \\
1.0 & $\cdots$ & -0.40 & -1.05 & 0.20 & 0.52 & 12.24 & -2.71 & 5.33 & 0.29 & 8.36 & 5.39 & 0.00 \\
\hline
\textbf{SPGI} (776) & \\
0.0 & 0.01 & -2.23 & -1.67 & 0.19 & 0.81 & 58.20 & -2.57 & 1.70 & 0.22 & 31.71 & 1.08 & 0.07 \\
0.0 & 0.02 & -2.17 & -1.78 & 0.18 & 0.84 & 72.88 & -2.63 & 3.28 & 0.22 & 20.60 & 1.09 & 0.04 \\
0.0 & 0.03 & -2.11 & -2.12 & 0.18 & 0.85 & 76.04 & -2.56 & 3.84 & 0.22 & 18.87 & 1.11 & 0.02 \\
0.0 & 0.04 & -2.06 & -2.39 & 0.19 & 0.89 & 73.79 & -2.43 & 3.86 & 0.22 & 20.19 & 1.50 & 0.00 \\
0.0 & 0.05 & -2.02 & -3.20 & 0.23 & 0.93 & 50.09 & -2.23 & 3.61 & 0.21 & 23.74 & 1.49 & 0.00 \\
0.0 & 0.06 & -1.98 & -3.87 & 0.22 & 1.05 & 60.16 & -1.91 & 3.48 & 0.20 & 28.32 & 2.33 & 0.00 \\
0.0 & 0.07 & -1.85 & -13.84 & 0.28 & 1.07 & 94.21 & -1.01 & 2.34 & 0.17 & 47.37 & 2.47 & 0.00 \\
0.0 & 0.08 & -1.97 & -6.64 & 0.21 & 1.19 & 112.27 & 8.18 & 1.94 & 0.08 & 0.10 & 2.75 & 0.00 \\
0.0 & 0.09 & -1.84 & -7.30 & 0.20 & 1.18 & 134.13 & 7.76 & 3.81 & 0.08 & 0.13 & 3.46 & 0.00 \\
0.0 & 0.10 & -2.54 & 0.45 & 0.16 & 0.73 & 13.21 & 6.82 & 9.03 & 0.09 & 0.20 & 5.43 & 0.00 \\
0.1 & 0.01 & -2.13 & -1.74 & 0.18 & 0.78 & 65.75 & -2.58 & 1.95 & 0.22 & 29.59 & 1.03 & 0.30 \\
0.1 & 0.02 & -2.07 & -1.70 & 0.16 & 0.81 & 99.34 & -2.61 & 3.33 & 0.22 & 20.78 & 1.05 & 0.18 \\
0.1 & 0.03 & -2.00 & -1.84 & 0.15 & 0.82 & 113.31 & -2.54 & 3.85 & 0.22 & 19.42 & 1.08 & 0.05 \\
0.1 & 0.04 & -1.93 & -2.27 & 0.16 & 0.85 & 94.54 & -2.40 & 3.79 & 0.22 & 21.31 & 1.28 & 0.00 \\
0.1 & 0.05 & -1.89 & -2.89 & 0.20 & 0.88 & 65.41 & -2.18 & 3.49 & 0.21 & 25.57 & 1.43 & 0.00 \\
0.1 & 0.06 & -1.86 & -4.05 & 0.23 & 0.94 & 53.47 & -1.86 & 3.52 & 0.20 & 29.46 & 2.49 & 0.00 \\
0.1 & 0.07 & -1.85 & -12.52 & 0.41 & 0.87 & 29.13 & -0.88 & 2.66 & 0.17 & 44.65 & 2.71 & 0.00 \\
0.1 & 0.08 & -1.90 & -6.83 & 0.24 & 1.06 & 83.84 & 8.10 & 1.47 & 0.08 & 0.16 & 2.73 & 0.00 \\
0.1 & 0.09 & -2.01 & -1.61 & 0.21 & 1.15 & 38.41 & 6.54 & 6.19 & 0.09 & 0.35 & 3.53 & 0.00 \\
0.1 & 0.10 & -0.90 & -0.50 & 0.20 & 0.00 & 282.28 & -0.34 & -104.07 & 0.16 & 74.93 & 9.76 & 0.00 \\
0.2 & 0.01 & -1.99 & -0.99 & 0.09 & 0.79 & 287.28 & -2.57 & 2.10 & 0.22 & 28.99 & 0.99 & 0.58 \\
0.2 & 0.02 & -1.92 & -1.22 & 0.10 & 0.79 & 232.74 & -2.60 & 3.54 & 0.22 & 20.43 & 1.01 & 0.44 \\
0.2 & 0.03 & -1.84 & -1.56 & 0.12 & 0.78 & 172.19 & -2.52 & 3.92 & 0.22 & 19.91 & 1.07 & 0.10 \\
0.2 & 0.04 & -1.77 & -2.34 & 0.17 & 0.78 & 86.11 & -2.36 & 3.79 & 0.22 & 22.36 & 1.18 & 0.00 \\
0.2 & 0.05 & -1.71 & -2.62 & 0.18 & 0.82 & 76.05 & -2.14 & 3.52 & 0.21 & 26.80 & 1.28 & 0.00 \\
0.2 & 0.06 & -1.73 & -6.95 & 0.47 & 0.58 & 12.79 & -1.75 & 2.89 & 0.20 & 36.55 & 2.94 & 0.00 \\
0.2 & 0.07 & -1.79 & -13.78 & 0.59 & 0.59 & 12.27 & -0.66 & 2.47 & 0.17 & 48.86 & 3.40 & 0.00 \\
0.2 & 0.08 & -1.80 & -5.35 & 0.21 & 0.97 & 90.34 & 8.04 & 1.84 & 0.08 & 0.15 & 2.73 & 0.00 \\
0.2 & 0.09 & -1.69 & -2.15 & 0.16 & 0.95 & 86.56 & 6.90 & 7.87 & 0.09 & 0.19 & 3.85 & 0.00 \\
0.2 & 0.10 & -1.05 & -0.84 & 0.09 & 0.70 & 72.86 & 6.61 & 8.65 & 0.09 & 0.29 & 5.72 & 0.00 \\
0.3 & 0.01 & -1.83 & -1.16 & 0.10 & 0.73 & 227.90 & -2.59 & 2.59 & 0.22 & 25.97 & 1.01 & 0.40 \\
0.3 & 0.02 & -1.72 & -1.12 & 0.09 & 0.73 & 271.75 & -2.59 & 3.79 & 0.22 & 20.11 & 1.03 & 0.28 \\
0.3 & 0.03 & -1.62 & -1.21 & 0.09 & 0.74 & 212.61 & -2.49 & 4.02 & 0.22 & 20.41 & 1.12 & 0.01 \\
0.3 & 0.04 & -1.53 & -2.81 & 0.20 & 0.67 & 53.61 & -2.32 & 3.86 & 0.22 & 23.15 & 1.30 & 0.00 \\
0.3 & 0.05 & -1.43 & -2.29 & 0.16 & 0.74 & 69.41 & -2.08 & 3.46 & 0.21 & 28.79 & 1.56 & 0.00 \\
0.3 & 0.06 & -1.48 & -8.12 & 0.63 & 0.31 & 8.04 & -1.66 & 2.75 & 0.19 & 40.47 & 3.34 & 0.00 \\
0.3 & 0.07 & -1.69 & -11.97 & 0.58 & 0.53 & 12.59 & -0.30 & 3.25 & 0.16 & 40.79 & 5.03 & 0.00 \\
0.3 & 0.08 & -1.67 & -3.71 & 0.18 & 0.87 & 106.09 & 8.57 & 2.03 & 0.07 & 0.06 & 2.75 & 0.00 \\
0.3 & 0.09 & -1.34 & -1.22 & 0.11 & 0.77 & 94.78 & 6.90 & 5.76 & 0.09 & 0.27 & 3.70 & 0.00 \\
0.3 & 0.10 & 0.40 & -0.26 & 0.03 & 0.53 & 8.84 & 6.35 & 10.50 & 0.10 & 0.39 & 6.11 & 0.00 \\
0.4 & 0.01 & -1.59 & -0.93 & 0.07 & 0.67 & 287.88 & -2.62 & 3.31 & 0.23 & 22.36 & 1.03 & 0.30 \\
0.4 & 0.02 & -1.45 & -0.87 & 0.07 & 0.68 & 253.71 & -2.57 & 4.10 & 0.23 & 19.46 & 1.05 & 0.15 \\
0.4 & 0.03 & -1.29 & -0.94 & 0.08 & 0.69 & 168.06 & -2.45 & 4.13 & 0.22 & 20.49 & 1.24 & 0.00 \\
0.4 & 0.04 & -1.11 & -1.40 & 0.11 & 0.68 & 78.70 & -2.28 & 3.88 & 0.22 & 23.62 & 1.62 & 0.00 \\
0.4 & 0.05 & -0.96 & -2.51 & 0.20 & 0.61 & 34.54 & -2.01 & 3.50 & 0.21 & 29.44 & 1.82 & 0.00 \\
0.4 & 0.06 & -0.99 & -3.84 & 0.32 & 0.52 & 20.57 & -1.54 & 2.85 & 0.19 & 41.63 & 4.20 & 0.00 \\
0.4 & 0.07 & -1.69 & -2.68 & 0.12 & 0.88 & 204.57 & 6.40 & 0.65 & 0.09 & 2.27 & 5.93 & 0.00 \\
0.4 & 0.08 & -1.42 & -1.71 & 0.11 & 0.79 & 157.77 & 6.81 & 4.61 & 0.09 & 0.33 & 2.99 & 0.00 \\
0.4 & 0.09 & -0.24 & -0.30 & 0.03 & 0.58 & 41.45 & 6.23 & 9.07 & 0.09 & 0.44 & 3.95 & 0.00 \\
0.4 & 0.10 & 1.21 & -0.22 & 0.03 & 0.49 & 1.05 & 8.04 & 7.22 & 0.09 & 0.12 & 7.05 & 0.00 \\
\hline 
\end{tabular}
\end{table*}
\begin{table*}
\contcaption{}
\centering     
\setlength{\tabcolsep}{6pt}
\begin{tabular}{c c c c c c c c c c c c c }  
\hline\hline                        
$b_C$ & $Y_{\text{Fe}}$ & $\alpha_g$ & $\beta_g$ & $a_{t,g}$ & $a_{c,g}$ & $C_g$ & $\alpha_s$ & $\beta_s$ & $a_{t,s}$ & $C_s$ & $\chi^2_\nu$ & p-value\\
$10^{-5}$ &  &  &  & $\mu$m & $\mu$m & $10^{-16}$ &  &  & $\mu$m & $10^{-16}$ &  &  \\
\hline
\textbf{SPGI} (776)& \\
0.5 & 0.01 & -1.23 & -0.77 & 0.06 & 0.63 & 201.61 & -2.63 & 4.06 & 0.23 & 18.93 & 1.07 & 0.08 \\
0.5 & 0.02 & -1.05 & -0.68 & 0.06 & 0.65 & 156.76 & -2.55 & 4.34 & 0.23 & 18.07 & 1.13 & 0.01 \\
0.5 & 0.03 & -0.85 & -0.86 & 0.07 & 0.66 & 80.44 & -2.42 & 4.20 & 0.23 & 19.41 & 1.51 & 0.00 \\
0.5 & 0.04 & -0.59 & -1.91 & 0.15 & 0.61 & 35.64 & -2.23 & 3.98 & 0.23 & 22.24 & 1.81 & 0.00 \\
0.5 & 0.05 & -0.33 & -2.87 & 0.21 & 0.54 & 24.78 & -1.91 & 3.46 & 0.21 & 29.42 & 3.55 & 0.00 \\
0.5 & 0.06 & -1.40 & -1.94 & 0.25 & 0.00 & 184.14 & -1.46 & -63.80 & 0.20 & 52.16 & 5.91 & 0.00 \\
0.5 & 0.07 & -1.48 & -1.66 & 0.09 & 0.80 & 238.53 & 5.31 & 1.05 & 0.10 & 5.29 & 7.56 & 0.00 \\
0.5 & 0.08 & -0.83 & -0.49 & 0.05 & 0.64 & 110.81 & 6.28 & 8.17 & 0.09 & 0.39 & 3.48 & 0.00 \\
0.5 & 0.09 & 0.83 & -0.55 & 0.05 & 0.48 & 6.09 & 7.27 & 7.02 & 0.09 & 0.24 & 4.65 & 0.00 \\
0.5 & 0.10 & 0.84 & 0.24 & 0.04 & 0.43 & 0.31 & 8.17 & 8.10 & 0.10 & 0.13 & 8.76 & 0.00 \\
0.6 & 0.01 & -0.85 & -0.59 & 0.05 & 0.62 & 114.69 & -2.63 & 4.46 & 0.24 & 16.11 & 1.36 & 0.00 \\
0.6 & 0.02 & -0.68 & -0.65 & 0.06 & 0.64 & 71.76 & -2.53 & 4.49 & 0.24 & 16.04 & 1.54 & 0.00 \\
0.6 & 0.03 & -0.25 & -3.81 & 0.12 & 0.59 & 57.17 & -2.39 & 4.29 & 0.24 & 17.54 & 2.04 & 0.00 \\
0.6 & 0.04 & -0.01 & -2.57 & 0.15 & 0.57 & 24.69 & -2.22 & 4.98 & 0.24 & 17.25 & 2.52 & 0.00 \\
0.6 & 0.05 & 0.23 & -2.67 & 0.24 & 0.47 & 14.74 & -1.77 & 3.19 & 0.22 & 31.30 & 4.09 & 0.00 \\
0.6 & 0.06 & -1.05 & -3.19 & 0.25 & 0.00 & 295.03 & -2.32 & -78.83 & 0.28 & 62.96 & 7.13 & 0.00 \\
0.6 & 0.07 & -1.05 & -0.71 & 0.06 & 0.68 & 179.87 & 4.85 & 0.80 & 0.10 & 10.51 & 10.24 & 0.00 \\
0.6 & 0.08 & 0.35 & -0.20 & 0.02 & 0.52 & 10.91 & 6.31 & 8.97 & 0.10 & 0.45 & 4.68 & 0.00 \\
0.6 & 0.09 & 1.51 & -0.58 & 0.06 & 0.45 & 1.81 & 8.41 & 5.87 & 0.09 & 0.11 & 6.00 & 0.00 \\
0.6 & 0.10 & 1.94 & 0.02 & 0.13 & 0.39 & 0.51 & 6.21 & 13.60 & 0.12 & 0.57 & 11.29 & 0.00 \\
0.7 & 0.01 & -0.56 & -1.00 & 0.09 & 0.60 & 47.20 & -2.63 & 4.70 & 0.25 & 13.64 & 1.95 & 0.00 \\
0.7 & 0.02 & -0.40 & -1.43 & 0.13 & 0.59 & 27.51 & -2.52 & 4.70 & 0.26 & 13.80 & 2.22 & 0.00 \\
0.7 & 0.03 & 0.00 & 0.14 & 0.18 & 0.00 & 327.66 & -2.96 & 4.46 & 0.19 & 9.59 & 2.64 & 0.00 \\
0.7 & 0.04 & 0.86 & -3.02 & 0.23 & 0.45 & 10.71 & -2.08 & 3.73 & 0.24 & 21.33 & 4.40 & 0.00 \\
0.7 & 0.05 & -1.18 & -1.77 & 0.25 & 0.00 & 197.77 & -2.53 & -50.62 & 0.26 & 45.68 & 7.03 & 0.00 \\
0.7 & 0.06 & -1.31 & -2.57 & 0.28 & 0.00 & 159.46 & -0.74 & -94.18 & 0.18 & 70.51 & 8.80 & 0.00 \\
0.7 & 0.07 & -0.00 & -1.21 & 0.11 & 0.50 & 28.36 & 3.60 & 11.93 & 0.12 & 3.16 & 11.77 & 0.00 \\
0.7 & 0.08 & 0.75 & 0.02 & 0.12 & 0.41 & 2.20 & 4.27 & 14.28 & 0.12 & 1.88 & 6.35 & 0.00 \\
0.7 & 0.09 & 1.21 & 0.15 & 0.09 & 0.41 & 0.62 & 8.14 & 9.48 & 0.10 & 0.13 & 8.12 & 0.00 \\
0.7 & 0.10 & 2.34 & 0.18 & 0.16 & 0.36 & 0.34 & 5.99 & 14.86 & 0.13 & 0.68 & 14.53 & 0.00 \\
0.8 & 0.01 & -0.39 & -0.96 & 0.10 & 0.60 & 28.37 & -2.62 & 4.92 & 0.27 & 11.56 & 3.67 & 0.00 \\
0.8 & 0.02 & -0.14 & -1.87 & 0.14 & 0.59 & 21.39 & -2.51 & 4.88 & 0.27 & 11.99 & 4.12 & 0.00 \\
0.8 & 0.03 & 0.46 & -2.31 & 0.15 & 0.55 & 12.48 & -2.32 & 4.37 & 0.26 & 14.75 & 4.55 & 0.00 \\
0.8 & 0.04 & 2.74 & 0.65 & 0.20 & 0.35 & 0.16 & -1.99 & 3.71 & 0.24 & 20.92 & 7.36 & 0.00 \\
0.8 & 0.05 & -0.84 & -1.75 & 0.22 & 0.00 & 338.29 & -2.83 & -55.52 & 0.20 & 47.18 & 10.27 & 0.00 \\
0.8 & 0.06 & -0.52 & -0.40 & 0.06 & 0.56 & 56.61 & 3.60 & 12.77 & 0.11 & 2.81 & 10.95 & 0.00 \\
0.8 & 0.07 & -0.61 & 0.21 & 0.20 & 0.00 & 213.28 & 2.05 & -245.49 & 0.06 & 162.16 & 14.30 & 0.00 \\
0.8 & 0.08 & 1.64 & -0.29 & 0.08 & 0.43 & 1.15 & 7.30 & 9.97 & 0.11 & 0.25 & 8.12 & 0.00 \\
0.8 & 0.09 & 2.59 & -0.35 & 0.10 & 0.40 & 0.38 & 7.96 & 10.97 & 0.11 & 0.17 & 10.98 & 0.00 \\
0.8 & 0.10 & 4.14 & 1.35 & 0.46 & 0.11 & 1.65 & 9.19 & 9.52 & 0.12 & 0.09 & 18.39 & 0.00 \\
0.9 & 0.01 & -0.25 & -0.97 & 0.11 & 0.59 & 19.02 & -2.62 & 5.14 & 0.28 & 9.85 & 4.91 & 0.00 \\
0.9 & 0.02 & -0.07 & -1.94 & 0.14 & 0.60 & 18.35 & -2.50 & 5.12 & 0.28 & 10.38 & 5.49 & 0.00 \\
0.9 & 0.03 & 7.56 & -2.44 & 0.42 & 0.17 & 3.56 & -2.25 & 4.31 & 0.27 & 13.96 & 5.97 & 0.00 \\
0.9 & 0.04 & -1.10 & -1.34 & 0.25 & 0.00 & 181.19 & -2.62 & -41.89 & 0.26 & 40.61 & 9.66 & 0.00 \\
0.9 & 0.05 & -1.09 & -2.39 & 0.28 & 0.00 & 171.13 & -2.55 & -53.39 & 0.25 & 46.97 & 12.09 & 0.00 \\
0.9 & 0.06 & -0.99 & -1.34 & 0.25 & 0.00 & 192.37 & -0.66 & -96.04 & 0.17 & 71.72 & 12.97 & 0.00 \\
0.9 & 0.07 & 1.62 & -0.45 & 0.03 & 0.45 & 0.81 & 8.51 & 5.70 & 0.09 & 0.09 & 16.61 & 0.00 \\
0.9 & 0.08 & -0.97 & 0.68 & 0.25 & 0.00 & 69.10 & -2.56 & -141.93 & 0.10 & 15.26 & 19.27 & 0.00 \\
0.9 & 0.09 & 2.68 & 0.03 & 0.18 & 0.35 & 0.39 & 7.00 & 13.77 & 0.13 & 0.38 & 14.64 & 0.00 \\
0.9 & 0.10 & 2.25 & 0.33 & 0.13 & 0.41 & 0.11 & 8.45 & 13.37 & 0.13 & 0.16 & 23.22 & 0.00 \\
1.0 & 0.01 & -0.22 & -0.71 & 0.11 & 0.60 & 12.98 & -2.62 & 5.29 & 0.29 & 8.55 & 6.50 & 0.00 \\
1.0 & 0.02 & -0.05 & -0.30 & 0.13 & 0.60 & 5.05 & -2.48 & 5.34 & 0.28 & 9.18 & 7.21 & 0.00 \\
1.0 & 0.03 & 2.55 & 0.57 & 0.15 & 0.41 & 0.06 & -2.18 & 4.18 & 0.27 & 14.14 & 7.72 & 0.00 \\
1.0 & 0.04 & -0.86 & -2.98 & 0.25 & 0.00 & 319.57 & -2.92 & -44.83 & 0.19 & 40.33 & 11.01 & 0.00 \\
1.0 & 0.05 & 1.09 & -1.63 & 0.14 & 0.46 & 7.44 & -0.01 & 16.97 & 0.17 & 9.25 & 12.91 & 0.00 \\
1.0 & 0.06 & -0.49 & 0.18 & 0.20 & 0.00 & 229.49 & 1.53 & -217.20 & 0.08 & 145.44 & 16.84 & 0.00 \\
1.0 & 0.07 & -0.93 & 0.03 & 0.25 & 0.00 & 100.30 & 3.33 & -156.35 & 0.88 & 20.31 & 20.27 & 0.00 \\
1.0 & 0.08 & 3.30 & -0.47 & 0.14 & 0.36 & 0.42 & 8.46 & 9.81 & 0.11 & 0.14 & 23.14 & 0.00 \\
1.0 & 0.09 & 3.90 & -0.41 & 0.24 & 0.31 & 0.61 & 8.71 & 11.01 & 0.13 & 0.14 & 19.00 & 0.00 \\
1.0 & 0.10 & 0.02 & 1.40 & 0.24 & 0.01 & 65.29 & 0.08 & -78.74 & 0.12 & 58.58 & 37.21 & 0.00 \\
\hline 
\end{tabular}
\end{table*}


\bsp	
\label{lastpage}
\end{document}